\documentclass[preprint,aps,prd,floatfix,superscriptaddress,preprintnumbers]{revtex4}
\usepackage{graphicx}
\def \s{\sqrt{2}}
\def \st{\sqrt{3}}
\def \sx{\sqrt{6}}

\def\be{\begin{equation}}
\def\ee{\end{equation}}
\def\bea{\begin{eqnarray}}
\def\eea{\end{eqnarray}}
\def\bean{\begin{eqnarray*}}
\def\eean{\end{eqnarray*}}
\def\bary{\begin{array}}
\def\eary{\end{array}}

\def\bit{\begin{itemize}}
\def\eit{\end{itemize}}
\def\bwt{\begin{widetext}}
\def\ewt{\end{widetext}}
\def\half{\frac{1}{2}}

\def\ol{\overline}

\begin{document}

\preprint{BNL-HET-06/10}

\title{DETERMINATION of $\gamma$ FROM CHARMLESS $B \to M_1 M_2$ DECAYS USING U-SPIN}

\author{Amarjit Soni}
\email[e-mail: ]{soni@quark.phy.bnl.gov}
\affiliation{High Energy Theory Group, 
Brookhaven National Laboratory, Upton, NY 11973}
\author{Denis A.~Suprun}
\email[e-mail: ]{suprun@quark.phy.bnl.gov}
\affiliation{High Energy Theory Group, 
Brookhaven National Laboratory, Upton, NY 11973}

\date{\today}

\begin{abstract}

%DS Hadronic $B$ decays to a pair of charmless mesons are analyzed within the context of U-spin, a subgroup of $SU(3)$ flavor symmetry. The U-spin based approach is complementary to the $SU(3)$-based quark-diagrammatic approach but it also has some important advantages. One of them is the higher precision of the U-spin symmetry. More importantly, no assumptions on the negligible size of exchange and annihilation amplitudes need to be made. The theoretical uncertainty of this method is expected to be small. Acceptable U-spin fits to the current $V^0 P^+$ and $P^0 P^+$ data allow the value of the weak phase $\gamma$ to be extracted from the fits: $\gamma=\left(54^{+12}_{-11}\right)^{\circ}$. We also identify modes ($\phi \pi^+$ and $\ol K^{*0} K^+$) where smaller uncertainties on their branching ratios would lead to a significantly more precise extraction of $\gamma$. We encourage experimental groups at BaBar and Belle to place stricter upper limits on these two modes than those that are currently available. 

%DS Neutral $B^0, B_s \to P^- P^+$ decays can also be used for this purpose. The fit to neutral $P^- P^+$ decays has two ambiguous solutions at $\gamma=\left(39\pm5\right)^{\circ}$ and $\gamma=\left(84^{+8}_{-12}\right)^{\circ}$. 

In our previous paper we applied U-spin symmetry to charmless hadronic $B^{\pm} \to M^0 M^{\pm}$ decays for the purpose of precise extraction of the unitarity angle $\gamma$. In this paper we extend our approach to neutral $B^0$ and $ B_s \to M_1 M_2$ decays.  A very important feature of this method is that no assumptions regarding relative sizes of topological decay amplitudes need to be made. As a result, this method avoids an uncontrollable theoretical uncertainty that is often related to the neglect of some topological diagrams (e.g., exchange and annihilation graphs) in quark-diagrammatic approaches. In charged $B^{\pm}$ decays, each of the four data sets, $P^0 P^{\pm}$, $P^0 V^{\pm}$, $V^0 P^{\pm}$ and $V^0 V^{\pm}$, with $P \equiv$ pseudoscalar and $V \equiv$ vector, can be used to obtain a value of $\gamma$. Among neutral decays, only experimental data in the $B^0, B_s \to P^- P^+$ subsector is sufficient for a U-spin fit. Application of the U-spin approach to the current charged and neutral $B$ decay data yields: $\gamma=\left(80^{+6}_{-8}\right)^{\circ}$. 
%DS TO BE UPDATED: We emphasize that improved measurements of $\phi \pi^{\pm}$ and $\ol K^{*0} K^{\pm}$ branching ratios would lead to appreciably better extraction of $\gamma$. 
In this method, which is completely data driven, in a few years we should be able to obtain a model independent determination of $\gamma$ with an accuracy of O(few degrees). 

\end{abstract}

%DS add if needed
%\pacs{13.25.Hw, 14.40.Nd, 11.30.Er, 11.30.Hv}
%11.30.Er  Charge conjugation, parity, time reversal, and other discrete
%          symmetries
%11.30.Hv  Flavor symmetries
%13.25.Hw  Decays of bottom mesons
%14.40.Nd  Bottom mesons

\maketitle

% Section I
\section{INTRODUCTION \label{sec:intro}}

Precise determinations of the angles of the unitarity
triangle (UT) remains an important but difficult goal in Particle Physics.
Though methods for direct determinations of all the angles are
now known, we are still quite far away from having large enough sample
of $B$'s that are needed~\cite{schune_ichep05}.
The main challenge in extracting the angles
from the data is of course that weak decays  take place in the
presence of strong interactions (i.e. QCD) which in this energy
regime has important, non-perturbative effects. Fortunately, QCD
respects flavor symmetries. Use of these symmetries presents
an important avenue to extract results, though often at the expense
of some accuracy. In the context of the angle
$\gamma$ of the UT, in fact SU(3) flavor symmetry has already been successfully employed~\cite{Chiang:2004nm,Chiang:2003pm,Suprun:2006}. Also, isospin symmetry can potentially be used for theoretically precise $\gamma$ extraction from three body charmless modes~\cite{Ciuchini:2006kv,Ciuchini:2006st,Gronau:2006qn}.
In this paper we show that U-spin can be used 
for determination of  $\gamma$ from charmless $B^{\pm}$, $B^0$ and $B_s$ decays. 

Previous studies~\cite{Chiang:2004nm,Chiang:2003pm,Suprun:2006} 
have explored $B$ meson decays to a pair of charmless pseudoscalar mesons ($PP$) or to a vector and pseudoscalar meson ($VP$) in the context of quark-diagrammatic approach and flavor SU(3) symmetry. Symmetry breaking was taken into account in tree amplitudes through ratios of decay constants; otherwise the exact SU(3) symmetry was assumed. Good separate fits to $PP$ and $VP$ data were obtained with tree ($T$),  color-suppressed ($C$), penguin ($P$ and $P_{tu}$), and electroweak penguin ($P_{EW}$) amplitudes. Other diagrams (exchange, annihilation, penguin annihilation) were assumed to be small and were neglected. Values of the weak phase $\gamma$ were extracted from the fits. They were found to be consistent with the current direct and indirect bounds on the CKM factor $\gamma$~\cite{angles}.

The quark-diagrammatic approach has two weak points. Firstly, the approach neglects exchange and annihilation contributions which some argue to be significant~\cite{Arnesen:2006vb,Kagan:2004uw,Mantry:2003uz}. Secondly, the extent at which flavor SU(3) symmetry is broken in $B$ decays cannot be accurately estimated within this model-independent approach. The intrinsic systematic uncertainty in $\gamma$ that is due to $SU(3)$ breaking effects is not completely under control and may happen to be substantial. 

These drawbacks motivate our current study of an alternative model-independent approach. $B$ meson decays can be explored within the framework of U-spin. There are substantial differences between U-spin multiplet approach and other phenomenological methods, such as SU(3) based approach, of understanding the current $B$ decay data.

\begin{itemize}

\item The significant advantage of U-spin {\it multiplet} method over SU(3) fits is that it makes fewer assumptions. In particular, U-spin method does not use quark diagrammatic {\it topological} approach at all. As a result, no assumptions about the relative sizes of various contributing topological diagrams are being made and no amplitude need be neglected~\cite{Fleischer:1999pa}.
The annihilation and exchange amplitudes that are usually neglected in SU(3) analyses~\cite{Chiang:2004nm,Chiang:2003pm} are formally of non-leading order and appear only at $O(1/m_b)$. 
However, $1/m_b$ corrections are notoriously difficult to reliably estimate; the $b$-quark mass ($\sim4.5$~GeV) is not so large compared to $\Lambda_{QCD}$ that such (formally) non-leading terms are necessarily negligible. Several models~\cite{Beneke:2001ev,Keum:2002vi,Cheng:2005bg,Kagan:2004uw} make highly varied estimates of these $1/m_b$ corrections that may appear with large chirally enhanced coefficients. 
However, in practice SU(3) fits~\cite{Chiang:2004nm,Chiang:2003pm} assumed that certain topologies give negligible contributions to limit the number of employed fit parameters for the purpose of fit stability. This assumption introduces into them a  model-dependent theoretical uncertainty.

%DS \item U-spin {\it multiplet} method has the significant advantage that, unlike SU(3) fits to charmless $B$ decays, quark diagrammatic {\it topological} approach is not invoked at all. Thus, we do not need to make any assumptions about the relative sizes of various contributing topological diagrams and so no amplitude need be neglected~\cite{Fleischer:1999pa}. The annihilation and exchange amplitudes that are usually neglected in SU(3) analyses~\cite{Chiang:2004nm,Chiang:2003pm} are formally of non-leading order and appear only at $O(1/m_b)$. However, $1/m_b$ corrections are notoriously difficult to reliably estimate; the $b$-quark mass ($\sim4.5$~GeV) is not so large compared to $\Lambda_{QCD}$ that such (formally) non-leading terms are necessarily negligible. Indeed, estimates of these $1/m_b$ corrections are highly model dependent~\cite{Beneke:2001ev,Keum:2002vi,Cheng:2005bg,Kagan:2004uw}. Furthermore, some of these corrections can appear with surprisingly large coefficients as they may be chirally enhanced. However, in practice SU(3) fits~\cite{Chiang:2004nm,Chiang:2003pm} assumed that certain topologies give negligible contributions to limit the number of employed fit parameters for the purpose of fit stability. This assumption introduces into them a  model-dependent theoretical uncertainty.

%DS deleted Consequently, the assumption of neglecting certain topologies that SU(3) fits~\cite{Chiang:2004nm,Chiang:2003pm} make use of introduces into them a  model-dependent theoretical uncertainty.

\item It is important to emphasize that the presence of flavor symmetry (SU(3) or U-spin) breaking effects does not necessarily translate into large  uncertainty in determination of $\gamma$. For instance, SU(3) breaking effects of about 20\% that are related to the ratio of decay constants $f_K$ and $f_{\pi}$, only lead to a small ($2^{\circ}$, or 3\%) theoretical uncertainty in determination of $\gamma$ from SU(3) fits~\cite{Suprun:2006}. 
Similarly, the theoretical error in $\gamma$ was found to be practically insensitive ($\lesssim1^{\circ}$~\cite{Chiang:2004nm}) to the uncertainty due to mixing in the definition of $\eta$ and $\eta'$ mesons. Since U-spin approach does not use graphical topologies, estimates of U-spin breaking effects on $\gamma$ extraction may be amenable to calculational frameworks such as QCD factorization, pQCD, soft collinear effective theory (SCET), or QCD sum rules~\cite{effective-theories}.

%DS \item It is important to emphasize that noticeable flavor symmetry breaking effects in decay amplitudes do not necessarily lead to large uncertainties in $\gamma$ extraction. For instance, SU(3) breaking effects of about 20\% that are related to the ratio of decay constants $f_K$ and $f_{\pi}$, only lead to a small ($2^{\circ}$, or 3\%) theoretical uncertainty in determination of $\gamma$ from SU(3) fits~\cite{Suprun:2005}. The effects of the $\eta-\eta'$ mixing on the theoretical error in $\gamma$ were also found to be small ($\lesssim1^{\circ}$)~\cite{Chiang:2004nm}. Since in the U-spin approach graphical topologies are not used, estimate of U-spin breaking effects on $\gamma$ extraction may be amenable to calculational frameworks such as QCD factorization, pQCD, soft collinear effective theory (SCET), or QCD sum rules~\cite{effective-theories}.

\item From group theory point of view U-spin is a flavor symmetry formally similar to isospin. While isospin symmetry breaking effects are smaller ($m_d/\Lambda_{QCD}$ vs.\ $m_s/\Lambda_{QCD}$), electroweak penguins require special treatment~\cite{Gronau:1998fn,Buras:1998rb} in isospin approach when it is applied to the problem of precise $\alpha$ extraction~\cite{Gronau:2004tm}. U-spin approach does not require any special modifications to include electroweak penguins. They do not break U-spin and are automatically included in effective U-spin amplitudes.

%DS \item From group theory point of view U-spin is a flavor symmetry formally similar to isospin. While isospin symmetry breaking effects are smaller ($m_d/\Lambda_{QCD}$ vs.\ $m_s/\Lambda_{QCD}$), electroweak penguins require special treatment~\cite{Gronau:1998fn,Buras:1998rb} in isospin approach when it is applied to the problem of precise $\alpha$ extraction~\cite{Gronau:2004tm}. The fact that the U-spin approach does not make use of graphical topologies, of course means that electroweak penguins are automatically fully contained within this approach.

\item Needless to say, the standard $B \to D K$ methods of direct $\gamma$ extraction are theoretically the cleanest (error of O(.1\%)~\cite{schune_ichep05}) and should ultimately provide the most accurate determination of $\gamma$. But this accuracy will only be attained after very large data samples become available, perhaps many years down the road. The U-spin approach, on the other hand, can provide a fairly accurate value of $\gamma$ (error of O(few percent)) with modest increase of luminosities. Furthermore, while the $B \to D K$ method does not involve penguins, the U-spin approach automatically includes all penguin contributions that are very important in charmless $B$ decays. The comparison of the values of $\gamma$ from the two methods provides a good test for new physics that is likely to reveal itself in loop diagrams.
\end{itemize}

In our previous paper~\cite{Soni:2005ah} we have shown that there are four separate sets of two-body decays of charged $B$'s each of which can give a value of $\gamma$.  In this paper we update results obtained from charged $B$ decays using the most recent experimental data and extend our approach to neutral $B^0$ and $B_s$ decays. We find that they allow even more precise determination of the weak phase, with an accuracy in the same ball park as other methods being used.
%DS restore if needed We identify modes ($\phi \pi^{\pm}$ and $\ol K^{*0} K^{\pm}$) whose improved experimental measurements should appreciably improve the accuracy on $\gamma$ with this method. 
In the era of
the current B-factories, with the planned luminosities
of a few $ab^{-1}$, the method should allow us to determine
$\gamma$ with an accuracy of a few degrees. Furthermore,
as better experimental information, at these luminosities,
becomes available for
all the relevant data sets, this  method should give an
understanding of its inherent systematic error.

%DS The most important advantage is that quark-diagrammatic approach is not used and no amplitudes are neglected. Instead, the effective Hamiltonian is written in the most general form. All physical decay amplitudes are expressed in terms of U-spin amplitudes that include all possible contributions from tree, penguin, exchange and annihilation diagrams that multiply the same weak factors. This inclusiveness of the U-spin approach  makes it particularly more attractive than $SU(3)$-symmetry-based methods.

%DS The magnitudes and relative strong phases of different U-spin amplitudes are extracted from $V^0 P^+$ data along with the value of the weak phase $\gamma$. The latter is of particular interest; the uncertainty in $\gamma$ extraction  will be reduced in the future when higher statistics on $B$ meson decays is accumulated.

We review U-spin notation and conventions in Section \ref{sec:uspin}.  We
derive physical decay amplitudes for charged $B$ decays in terms of U-spin amplitudes in Section \ref{sec:charged}. Section \ref{sec:data} reviews the current experimental data on charmless $B^+ \to PP, VP, VV$ decays. U-spin fit results are presented in Section \ref{sec:results}.
%DS while the potential for improvement is discussed in Section \ref{sec:future}. 
Neutral decays are discussed in Section \ref{sec:neutral} with particular attention to decays into two oppositely charged mesons in Sections \ref{sec:neutral-to-two-charged} and 
\ref{sec:neutral-to-two-charged exp data} and to decays into two neutral charmless mesons in Sections \ref{sec:neutral-to-neutral}, \ref{sec:neutral-to-neutral amplitudes}, 
\ref{sec:neutral-to-neutral exp data}, \ref{sec:neutral-to-neutral V0P0} and \ref{sec:neutral-to-neutral full V0P0}. Section \ref{sec:summary} concludes. Appendix A shows the current experimental data on branching ratios and $CP$ asymmetries for all $B$ and $B_s$ decay modes.

\section{U-spin \label{sec:uspin}}

Let us very briefly recapitulate some elementary aspects of 
U-spin~\cite{Chiang:2003pm,Chiang:2003rb,recents,Fleischer:1999pa}.    
Recall that the U-spin subgroup of $SU(3)$ is 
similar to the I-spin (isospin) subgroup except that the quark doublets with $U = 1/2, U_3 = \pm 1/2$ are
\be
\label{eqn:qks}
{\rm Quarks:}~~\left[ \begin{array}{c} |\half~~\half \rangle \\
                                      |\half -\! \half \rangle \end{array}
\right] = \left[ \begin{array}{c} |d \rangle \\ |s \rangle \end{array}
\right]~~,
\ee
\be
{\rm Antiquarks:}~~\left[ \begin{array}{c} |\half~~\half \rangle \\
                                      |\half -\!\half \rangle \end{array}
\right] = \left[ \begin{array}{c} |\bar s \rangle \\ -\!| \bar d \rangle
\end{array} \right]~~.
\ee
$B^+$ is a U-spin singlet, while $\pi^+ (\rho^+)$, $K^+ (K^{*+})$ and their antiparticles belong to U-spin doublets,
\be
|0~~0 \rangle = |B^+\rangle = |u\bar b\rangle~~,
\ee
\be
\label{eqn:doublet}
\left[ \begin{array}{c} |\half~~\half \rangle \\
                                      |\half -\!\half \rangle \end{array}
\right] = \left[ \begin{array}{rcl} |u\bar s \rangle & = & |K^+~(K^{*+})\rangle\\ 
-\!| u\bar d \rangle & = & -|\pi^+~(\rho^+)\rangle
\end{array} \right]~~,
\ee
\be
\label{eqn:doublet2}
\left[ \begin{array}{c} |\half~~\half \rangle \\
                     |\half -\!\half \rangle \end{array} \right] = 
\left[ \begin{array}{rcl} |\bar u d \rangle & = & -|\pi^-~(\rho^-) \rangle\\ 
\!| \bar u s \rangle & = & -|K^-~(K^{*-})\rangle
\end{array} \right]~~.
\ee

Nonstrange neutral mesons belong either to a U-spin triplet or a U-spin
singlet.  We take $\pi^0 \equiv (d \bar d - u \bar u)/\sqrt2$, $\eta_8 \equiv (2 s \bar s - u \bar u - d \bar d)/\sqrt{6}$ and $\eta_1 \equiv (u \bar u + d \bar d + s \bar s)/\sqrt{3}$. The U-spin triplet is
\be
\label{eqn:tripletP}
\left[ \begin{array}{c} |1~~1 \rangle \\ |1~~0 \rangle \\ |1 -\!1 \rangle
\end{array} \right] = \left[ \begin{array}{c}
|d \bar s \rangle = |K^0 \rangle \\
\frac{1}{\sqrt{2}} | s \bar s - d \bar d \rangle = \frac{\sqrt{3}}{2} |\eta_8 \rangle - \frac{1}{2} |\pi^0 \rangle\\
-\! |s \bar d \rangle = -\! |\ol K^0 \rangle  \end{array} \right]~~,
\ee
and the corresponding singlet residing in the pseudoscalar meson octet is
\be
|0~~0 \rangle_8 \equiv \frac{1}{\sqrt{6}} | s \bar s + d \bar d - 2u \bar u \rangle = \frac{1}{2} |\eta_8 \rangle + \frac{\sqrt{3}}{2} |\pi^0 
\rangle ~~.
\ee 
In addition, there is another U-spin singlet which does not belong to the meson octet. Besides being a U-spin singlet, it is also an SU(3) singlet: 
\be
|0~~0 \rangle_1 \equiv \frac{1}{\sqrt{3}} | u \bar u + d \bar d + s \bar s \rangle = |\eta_1 \rangle ~~.
\ee 
The physical $\eta$ and $\eta'$ are mixtures of the octet and singlet. A straightforward calculation casts $\pi^0$, $\eta$ and $\eta'$ in terms of linear combinations of the U-spin multiplets:
%
%$$
\bea
\label{eqn:P-neutrals}
\begin{array}{llrll}
& & \pi^0 \quad  & = &  \quad  -\frac12 |1~~0 \rangle + \frac{\sqrt3}{2} |0~~0 \rangle_8~~,\\
\eta \quad & = & \frac{2\sqrt 2}{3} \eta_8 - \frac{1}{3} \eta_1 & = &  
\quad  \sqrt{\frac23} |1~~0 \rangle + \frac{\sqrt2}{3} |0~~0 \rangle_8 - 
\frac13 |0~~0 \rangle_1~~,\\
\eta' \quad & = & \frac{2\sqrt 2}{3} \eta_1 + \frac{1}{3} \eta_8 & = &
 \quad  \frac{1}{2\sqrt3} |1~~0 \rangle + \frac16 |0~~0 \rangle_8 + 
\frac{2\sqrt 2}{3} |0~~0 \rangle_1~~.
\end{array}
\eea
%$$

Similarly, the U-spin triplet in the vector meson octet is
\be
\label{eqn:tripletV}
\left[ \begin{array}{c} |1~~1 \rangle \\ |1~~0 \rangle \\ |1 -\!1 \rangle
\end{array} \right] = \left[ \begin{array}{c}
|d \bar s \rangle = |K^{*0} \rangle \\
\frac{1}{\sqrt{2}} | s \bar s - d \bar d \rangle = 
\frac{1}{\s} |\phi \rangle - \frac{1}{2} |\rho^0\rangle - 
\frac{1}{2}|\omega\rangle \\
-\! |s \bar d \rangle  = -\! |\ol K^{*0} \rangle \end{array} \right]~~,
\ee
while the corresponding singlet in the vector meson octet is
\be
\label{eqn:singletV}
|0~~0 \rangle_8 = \frac{1}{\sqrt{6}} | s \bar s + 
d \bar d - 2u \bar u \rangle = \frac{1}{\sx} |\phi \rangle + \frac{\sqrt{3}}{2} |\rho^0 \rangle - \frac{1}{2\st}|\omega \rangle ~~.
\ee 
The SU(3) singlet in the vector meson sector is given by
\be
|0~0 \rangle_1 = \frac{1}{\sqrt6} | u \bar u + d \bar d + s \bar s \rangle = (|\phi\rangle + \sqrt{2}|\omega\rangle)/\st.
\ee
Thus, the multiplet decompositions of $\rho^0$, $\omega$ and $\phi$ can be determined to be
%$$
\bea
\label{eqn:V-neutrals}
\begin{array}{lll}
\rho^0 \quad  & = &  \quad -\frac12 |1~~0 \rangle + \frac{\sqrt3}{2} |0~~0 \rangle_8~~,\\
\omega \quad & = &  \quad -\frac12 |1~~0 \rangle - \frac{\sqrt3}{6} |0~~0 \rangle_8 + \sqrt\frac23 |0~~0 \rangle_1~~,\\
\phi \quad & = &  \quad \frac{1}{\sqrt 2} |1~~0 \rangle + 
\frac{1}{\sqrt 6} |0~~0 \rangle_8 + \frac{1}{\sqrt 3} |0~~0 \rangle_1~~.
\end{array}
\eea
%$$

One may decompose the strangeness-conserving $\Delta S=0$ and strangeness-changing $|\Delta S|=1$ effective Hamiltonians
into members of {\em the same} two U-spin doublets multiplying given 
CKM factors.
%DS \cite{Uspin}.
For practical purposes, it is convenient to use a convention in which
the CKM factors involve the $u$ and $c$ quarks: 
%DS \cite{Uspin}.
%
\bea
\label{Hd}
\Delta S=0: \quad {\cal H}_{\rm eff}^{\bar b\to\bar d} & = & V^*_{ub}V_{ud}O^u_d + 
V^*_{cb}V_{cd}O^c_d~~,\\
\label{Hs}
|\Delta S|=1: \quad {\cal H}_{\rm eff}^{\bar b\to\bar s} & = & V^*_{ub}V_{us}O^u_s + 
V^*_{cb}V_{cs}O^c_s~~.
\eea
%
%DS add how EWPs break SU(3) symmetry but not U-spin
The assumption of U-spin symmetry implies that U-spin doublet 
operators $O^u_d$ and $O^u_s$ are identical, as well as 
the $O^c_d$ and $O^c_s$ operators. The subscripts $d$ and $s$ 
may be omitted. Hadronic matrix elements of these two 
operators, $O^u$ and $O^c$, will be denoted $A^u$ and $A^c$ and will 
be referred to as ``u-like" and ``c-like" amplitudes~\cite{tree_foot}, 
where the latter includes electroweak penguin contributions.  Note that these amplitudes multiply different CKM factors in $|\Delta S|=1$ and $\Delta S=0$ processes.

%DS old
% In isospin analysis of $B$ decays~\cite{Gronau:1990ka} the effective Hamiltonian transforms as $\Delta I=\frac12$ in $\bar b \to \bar d q \bar q$ penguin diagrams (the $q\bar q$ pair that couples to the mediating gluon, photon, or $Z$ boson must be in the $I=0$ state) and as either $\Delta I=\frac12$ or $\Delta I=\frac32$ in  $\bar b \to \bar u u \bar d$ tree diagrams ($u\bar d$, that couples to $W^+$ is in the $|1~~1\rangle$ state, combines with $\bar u=-|\frac12 ~~ -\frac12\rangle$ to form $\Delta I=\frac12$ and $\Delta I=\frac32$ states). 

%DS old
% In U-spin analysis the effective Hamiltonian always transforms as a single U-spin state. For instance, in  $\bar b \to \bar u u \bar d$ tree diagrams  $u \bar d$, which is a U-spin doublet, couples with a U-spin singlet $\bar u$ to form a $\Delta U_3 = -\half$ state. In U-spin approach there is no significant difference between tree and penguin diagrams but only between strangeness-conserving and strangeness-changing processes. The $|\Delta S| = 1$ effective Hamiltonian always transforms like a $\bar s \sim |\half \half \rangle$, that is, like a $\Delta U_3 = \half$ component of a U-spin doublet, while the $\Delta S = 0$ Hamiltonian always transforms like a $\bar d \sim -|\half -\half \rangle$, i.e.\, like a $\Delta U_3 = -\half$ component of a U-spin doublet. 

In isospin analysis of $B$ decays~\cite{Gronau:1990ka} the effective Hamiltonian transforms as either $\Delta I=\frac12$ or $\Delta I=\frac32$. 
While electroweak penguins violate isospin due to the charge difference between $u$ and $d$ quarks, they do not violate U-spin. In U-spin analysis 
%DS the effective Hamiltonian always transforms as a U-spin doublet, $\Delta U = \half$. The 
$|\Delta S| = 1$ effective Hamiltonian ${\cal H}_{\rm eff}^{\bar b\to\bar s}$ transforms like a $\bar s \sim |\half \half \rangle$, that is, like a $\Delta U_3 = \half$ component of a U-spin doublet $\Delta U = \half$. At the same time $\Delta S = 0$ Hamiltonian  ${\cal H}_{\rm eff}^{\bar b\to\bar d}$ transforms like a $\bar d \sim -|\half -\half \rangle$, i.e.\, like a $\Delta U_3 = -\half$ component of the U-spin doublet. 

There are three topological diagrams that may contribute to charged $B$ decays: tree, penguin (QCD and electroweak), and annihilation. In U-spin approach the effective Hamiltonian of any of these decay types always transforms as a U-spin doublet, $\Delta U = \half$. There is no principal  difference between tree, penguin and annihilation contributions but only between strangeness-conserving and strangeness-changing processes. This makes U-spin a particularly convenient approach that allows the complete description of charged $B$ decays without making any assumptions on the size of individual topological diagrams and without neglecting any of them, including annihilation. 

%DS While the SU(3) based approach~\cite{Chiang:2004nm,Chiang:2003pm} does not inherently require ignoring annihilation, exchange and penguin annihilation contributions, in practice one has to do that to limit the number of parameters and keep SU(3) fits stable. This advantage of the U-spin approach makes it particularly appealing; in the long run, it should significantly reduce theoretical uncertainties associated with this method.

\section{Charged $B$ decays \label{sec:charged}}

Since the initial $B^+$ meson is a U-spin singlet and the effective Hamiltonian always transforms as a U-spin doublet, the final $M^0 M^+$ states must be U-spin doublets. They can be formed in three different ways.

While the charmless charged meson $M^+$ can only belong to the  doublet~(\ref{eqn:doublet}), the neutral meson $M^0$ can be a linear combination of three different multiplets. Four neutral $K$ mesons ($K^0$, $K^{*0}$, and their antiparticles) contain only the triplet contribution, either $|1~~1\rangle$ or $|1~~-1\rangle$. The other neutral mesons~(\ref{eqn:P-neutrals}), (\ref{eqn:V-neutrals}) are linear combinations of the  $|1~~0\rangle$ triplet state, the U-spin singlet $|0~~0\rangle_8$ state and the SU(3) singlet $|0~~0\rangle_1$ state. As a result, any 
strangeness-conserving $\Delta S = 0$ $B^+ \to M^0 M^+$ decay amplitude can be expressed in terms of three amplitudes: $A^d_1$, $A^d_{0_8}$, $A^d_{0_1}$ that were denoted $A^d_1$, $A^d_0$, $B^d_0$ in~\cite{Chiang:2003pm}. They correspond to the U-spin triplet, U-spin singlet, and SU(3) singlet contributions into the decay amplitude, respectively. Each of these three amplitudes consists of a ``u-like" and a ``c-like" part, for instance, $A^d_1  =  V^*_{ub}V_{ud}A^u_1 + V^*_{cb}V_{cd}A^c_1$. Similarly, any  strangeness-changing $|\Delta S|=1$ decay amplitude can be written in terms of three other amplitudes: $A^s_1$, $A^s_{0_8}$, $A^s_{0_1}$. The assumption of U-spin symmetry implies that the difference between $A^d_1$ and $A^s_1$ comes only through the difference in the weak couplings, that is, $A^s_1  =  V^*_{ub}V_{us}A^u_1 + V^*_{cb}V_{cs}A^c_1$. Thus, the complete amplitudes for U-spin final states are given by
\bea
\label{eqn:Ad}
\Delta S=0: \quad A^d_{1, 0_8, 0_1} & = & V^*_{ub}V_{ud}A^u_{1, 0_8, 0_1} + V^*_{cb}V_{cd}A^c_{1, 0_8, 0_1},\\
\label{eqn:As}
|\Delta S|=1: \quad A^s_{1, 0_8, 0_1} & = & 
V^*_{ub}V_{us}A^u_{1, 0_8, 0_1} + V^*_{cb}V_{cs}A^c_{1, 0_8, 0_1}.
\eea

Consider, for instance, the $B^+ \to \pi^0 K^+$ decay. The initial meson $B^+$ is a U-spin singlet $|0~~0 \rangle$ that is affected by the transformation ($|\Delta S| = 1$ Hamiltonian) which is a U-spin doublet $|\half~~\half \rangle$. The final state is a combination of 
$\pi^0 =  \frac{\sqrt3}{2} |0~~0 \rangle_8  -\frac12 |1~~0 \rangle$
and the $|\half~~\half \rangle$ doublet $K^+$. The $U=\frac32$ final state cannot contribute to this decay process, the only contribution to the decay amplitude comes from the $U=\frac12$ final state. Using the Clebsch-Gordan coefficients, we determine that $A(\pi^0 K^+)$ is proportional to $\st A^s_{0_8} + \frac{1}{\st} A^s_1$. Similarly, one can calculate U-spin expressions for all $M^0 M^+$ decay amplitudes.

Following the conventions of~\cite{Chiang:2003pm}, we absorb the $\frac{1}{2\sqrt3}$ factor into the definition of $A^{d,s}_1$, the $\frac{\sqrt3}{2}$ factor into the definition of $A^{d,s}_0$ and the $\frac{1}{\sqrt3}$ factor into the definition of $B^{d,s}_0$. Then we find that physical decay amplitudes for $V^0 P^+$ and $V^0 V^+$ modes may be decomposed into U-spin amplitudes,
%$$
\bea
\begin{array}{llcc}
\label{eqn:KK}
A(\ol K^{*0} K^+),\quad & A(\ol K^{*0} K^{*+}) & = & \quad - 2\s A^d_1~~,\\
A(\rho^0\pi^+), & A(\rho^0\rho^+) & = & \quad A^d_{0_8} - A^d_1~~,\\
A(\omega\pi^+), & A(\omega\rho^+) & = & \quad -\frac13A^d_{0_8} - A^d_1 + \frac{\s}{3} A^d_{0_1}~~,\\
\label{eqn:phipi}
A(\phi \pi^+), & A(\phi \rho^+) & = & \quad \frac{\s}{3} A^d_{0_8} + \s A^d_1 + \frac13 A^d_{0_1}~~,\\
A(K^{*0}\pi^+), & A(K^{*0}\rho^+) & = & \quad -2\s A^s_1~~,\\
A(\rho^0 K^+), & A(\rho^0 K^{*+}) & = & \quad A^s_{0_8} + A^s_1~~,\\
A(\omega K^+), & A(\omega K^{*+}) & = & \quad -\frac13A^s_{0_8} + A^s_1 +\frac{\s}{3} A^s_{0_1}~~,\\
A(\phi K^+), & A(\phi K^{*+}) & = & \quad \frac{\s}{3} A^s_{0_8} -\s A^s_1 + \frac13 A^s_{0_1}~~,
\end{array}
\eea
%$$
where $A_1$, $A^d_{0_8}$ and $A^d_{0_1}$ correspond to final states with vector mesons $V^0$ in the U-spin triplet, in the octet U-spin singlet and in the SU(3) singlet, respectively. Naturally, the formulae for related $V^0 P^+$ and 
$V^0 V^+$ decay modes are the same, as seen in the above relations. However, the actual values for each of the U-spin amplitudes are constant only within each of the two subsets. They accept different values in $V^0 P^+$ and $V^0 V^+$ subsets.

Thus, eight $V^0 P^+$ decays are described by 12 parameters: six U-spin amplitudes $|A^u_{1, 0_8, 0_1}|$ and $|A^c_{1, 0_8, 0_1}|$, five relative strong phases between them and the weak phase $\gamma$. The same statement is separately valid for eight $V^0 V^+$ modes, too.

In the same way one can decompose physical amplitudes for $P^0 P^+$ and 
$P^0 V^+$ decay modes into U-spin amplitudes. We follow the conventions of~\cite{Chiang:2003rb} and absorb the $\frac{1}{2}$ factor into the definitions of $A^{d,s}_{1, 0_8, 0_1}$. Then we derive:
%$$
\bea
\begin{array}{llcc}
\label{eqn:Uspin}
A(\ol K^0 K^+),\quad & A(\ol K^0 K^{*+}) & = & \quad -\frac{4}{\sx}A^d_1~~,\\
A(\pi^0\pi^+), & A(\pi^0\rho^+) & = & \quad \st A^d_{0_8} - \frac{1}{\st}A^d_1~~,\\
A(\eta\pi^+), & A(\eta\rho^+) & = & \quad \frac{2\s}{3}A^d_{0_8}+\frac{2\s}{3}A^d_1 -\frac13 A^d_{0_1}~~,\\
A(\eta'\pi^+), & A(\eta'\rho^+) & = & \quad \frac13 A^d_{0_8}+\frac13 A^d_1+\frac{2\s}{3} A^d_{0_1}~~,\\
A(K^0\pi^+), & A(K^0\rho^+) & = & \quad -\frac{4}{\sx}A^s_1~~,\\
A(\pi^0 K^+), & A(\pi^0 K^{*+}) & = & \quad \st A^s_{0_8} + \frac{1}{\st} A^s_1~~,\\
A(\eta K^+), & A(\eta K^{*+}) & = & \quad \frac{2\s}{3}A^s_{0_8}-\frac{2\s}{3}A^s_1 -\frac13 A^s_{0_1}~~,\\
A(\eta' K^+), & A(\eta' K^{*+}) & = & \quad \frac13 A^s_{0_8}-\frac13 A^s_1+\frac{2\s}{3} A^s_{0_1}~~.
\end{array}
\eea
%$$

Just as the two subsets of $M^0 M^+$ that were considered before, $P^0 P^+$ and $P^0 V^+$ are also separately described by 12 parameters: six U-spin amplitudes $|A^u_{1, 0_8, 0_1}|$ and $|A^c_{1, 0_8, 0_1}|$, five relative strong phases between them and the weak phase $\gamma$.

All six U-spin amplitudes are essentially effective amplitudes. They may contain several topological amplitudes: trees, penguins, color-suppressed amplitudes, and annihilations (exchanges and penguin annihilations do not contribute to charged $B$ decays). For instance, $A^u_{0_1}$ is an amplitude that contributes to all $(\omega,\phi,\eta,\eta')M^+$ decays. It gets multiplied by the product of CKM factors $V^*_{ub}V_{u(d,s)}$. This means that $A^u_{0_1}$ accepts contributions from trees, color-suppressed diagrams, exchanges and $u$-quark mediated parts of QCD and electroweak penguins. The same can be said about the other two U-spin amplitudes with superscript $u$. On the other hand, the U-spin amplitudes with superscript $c$ only get contributions from $c$-quark mediated parts of QCD and electroweak penguins. The most important advantage of the U-spin approach is that one does not have to assume that annihilations, exchanges and penguin annihilations are negligible. While the SU(3) based approach~\cite{Chiang:2004nm,Chiang:2003pm} does not inherently require making these simplifying assumptions, in practice one has to do that to limit the number of parameters and keep SU(3) fits stable. This advantage of the U-spin approach makes it particularly appealing. It reduces theoretical uncertainties associated with this method.

\subsection{Review of the experimental data \label{sec:data}}

Charmless hadronic decays of the $B^+$ meson to the two-meson final state that contains vector $V$ or pseudoscalar $P$ mesons comprise four subsets: $P^0 P^+$, $V^0 V^+$, $V^0 P^+$, and $P^0 V^+$. Each of the subsets consists of eight decays, with all possible combinations of two charged mesons (e.g., $\pi^+$ and $K^+$ in the pseudoscalar octet) and four neutral ones (e.g., $K^{*0}$, $\rho$, $\omega$, and $\phi$ in the vector octet). Thus, there are altogether 16 relevant $B^{\pm}$ decays of each of the four types. Each of the subsets, again, is described by 12 parameters, namely, 6 U-spin amplitudes, 5 relative strong phases between them, and the weak phase $\gamma$ which is the only common parameter among four parameter sets. Thus, $\gamma$ can be separately determined from each subset. Alternatively, one can do the joint fit to determine the value of $\gamma$ that is most consistent with all four data sets simultaneously. Both avenues have been explored.   

All 8 $B^+ \to P^0 P^+$ decays have actually been observed and their branching ratios and $CP$ asymmetries have been measured, though, with the present statistics in most cases the errors are rather large.  This is especially so for the CP-asymmetries. In any case, with 16 data points and 12 fit parameters one can perform a fit and extract the preferred values for all parameters. 

In the other 3 subsets some modes have not yet been observed but upper limits on their branching ratios were reported. Needless to say, direct $CP$ asymmetries for these modes have not been determined yet. For some of these modes a central value and a large uncertainty are known. For the others, where only an upper limit at $90\%$ confidence level is reported, one can take central value as equal to 0 and approximately estimate the uncertainty by dividing the upper limit value by 2. For example, from ${\cal B}(B^+ \to \omega \rho^+)<16$ we crudely estimate that ${\cal B}(\omega \rho^+)=0.0\pm8.0$~\cite{units}. The data from upper limits helps in two ways. First of all, it provides additional data points, making a U-spin fit feasible. Second, it ensures that the resulting fit is consistent with the current upper limits.

In the case of $V^0 P^+$ decays, for instance, 6 out of 8 modes have been observed and provide 12 data points. The remaining two decays, $\ol K^{*0} K^+$ and $\phi \pi^+$, have not yet been observed. At present only the upper limits for these two modes are known: 
${\cal B}(B^+ \to \ol K^{*0} K^+)=0.0^{+1.3+0.6}_{-0.0-0.0} \; (<5.3)$~\cite{Jessop:2000bv} and 
${\cal B}(B^+ \to \phi \pi^+)=-0.04\pm0.17^{+0.03}_{-0.04} \; (<0.24)$~\cite{Aubert:2006nn}. 
From these measurements we can estimate that 
\bea
\label{eqn:KKdata}
{\cal B}(B^+ \to \ol K^{*0} K^+)=0.00^{+1.43}_{-0.00},\\  
\label{eqn:phipidata}
{\cal B}(B^+ \to \phi \pi^+)=-0.04\pm0.17. 
\eea
To make sure that the fit is consistent with the upper limits on the 
$\ol K^{*0} K^+$ and $\phi \pi^+$ branching ratios we add two more data points to the fit. Thus, the 12-parameter $V^0 P^+$ U-spin fit features 14 data points, making  $\gamma$ extraction possible.

Similarly, in the $V^0 V^+$ sector 5 modes have been detected and their $CP$ asymmetries measured, for the total of 10 data points. The other 3 modes have not yet been observed but the upper limits were reported, allowing estimates of their branching ratios. The total number of $V^0 V^+$ data points rises to 13.

The least is known about $P^0 V^+$ decays. Not even an upper limit is known for $\bar{K^0} K^{*+}$ branching ratio. However, first  measurements of $CP$ asymmetries in $\eta' \rho^+$ and $\eta' K^{*+}$ decays increased the total number of measured data points up to 13, allowing the 12-parameter U-spin fit 

%DS The least is known about $P^0 V^+$ decays. Not even an upper limit is known for $\bar{K^0} K^{*+}$. Of the remaining 7 decay modes only 4 have been detected, providing 8 data points. For the other three an estimate of the branching ratio can be made using current upper limits. Thus, there are only 11 data points and a reasonable 12 parameter U-spin fit cannot be performed. To avoid this problem, one can make a joint U-spin fit to two $M^0 M^+$ decay subsets, e.g.\ a fit to both $V^0 P^+$ and $P^0 V^+$ decays. With $\gamma$ being the only common parameter for both parameter sets, there are 11 completely free $P^0 V^+$ U-spin parameters (amplitudes and strong phases) that describe 11 $P^0 V^+$ data points. There is just enough data to make the joint fit work.

\subsection{Results \label{sec:results}}

Table~\ref{tab:fits} shows the results of the U-spin fits to four subsets of $M^+ M^0$ decays and their combinations. The top part of the table shows  fits to four individual subsets ($P^0 P^+$, $V^0 P^+$, $P^0 V^+$, $V^0 V^+$). The $P^0 P^+$ fit features a good $\chi^2=3.2/4$ and a relatively deep minimum at $\gamma=81^{\circ}$. The other three U-spin fits in the top part of the table produce very shallow minima, leaving $\gamma$ practically undetermined. One can make the conclusion that $V^0 P^+$, $P^0 V^+$ and $V^0 V^+$ data is not expected to significantly affect joint fits. 

\begin{table}[tp]
\begin{center}
\small
\caption{Results of the U-spin fits to various subsets of charmless $B^+ \to M^+ M^0$ decays.
\label{tab:fits}}
\begin{tabular}{clccc}
\hline
\hline
%\quad Fit~\quad\quad & Subset & Modes &\quad $\chi^2/dof$~\quad\quad & \quad\quad $\gamma$~\quad\quad\quad \\
Fit & Subset & Modes &$\chi^2/dof$ & $\gamma$\\
\hline
1.& $P^0 P^+$ & ~\quad
$\ol K^0 K^+$\, $\pi^0\pi^+$\,$\eta\pi^+$\,$\eta'\pi^+$\quad 
$ K^0 \pi^+$\,$\pi^0K^+$\,$\eta K^+$\,$\eta' K^+$\quad\quad  & 
3.24/4 & $\left(81^{+36}_{-18}\right)^{\circ}$  \\
%DS 2.& $V^0 P^+$ & ~\quad $\ol K^{*0} K^+$\, $\rho^0\pi^+$\,$\omega\pi^+$\,$\phi\pi^+$\quad $ K^{*0} \pi^+$\,$\rho^0K^+$\,$\omega K^+$\,$\phi K^+$\quad\quad  & 5.20/2 & $\left(19^{+12}_{-19}\right)^{\circ}$  \\
2.& $V^0 P^+$ & ~\quad
$\ol K^{*0} K^+$\, $\rho^0\pi^+$\,$\omega\pi^+$\,$\phi\pi^+$\quad 
$ K^{*0} \pi^+$\,$\rho^0K^+$\,$\omega K^+$\,$\phi K^+$\quad\quad   & 
1.80/2 & $\left(90\pm61\right)^{\circ}$  \\
3.& $P^0 V^+$ & ~\quad
$\ol K^0 K^{*+}$\, $\pi^0\rho^+$\,$\eta\rho^+$\,$\eta'\rho^+$\quad 
$ K^0 \rho^+$\,$\pi^0K^{*+}$\,$\eta K^{*+}$\,$\eta' K^{*+}$\quad\quad  
& 0.04/1 & $\left(47^{+133}_{-47}\right)^{\circ}$ \\ 
4.& $V^0 V^+$ & ~\quad
$\ol K^{*0} K^{*+}$\, $\rho^0\rho^+$\,$\omega\rho^+$\,$\phi\rho^+$\quad 
$ K^{*0} \rho^+$\,$\rho^0K^{*+}$\,$\omega K^{*+}$\,$\phi K^{*+}$\quad\quad
& 0.01/1 & $\left(23^{+157}_{-23}\right)^{\circ}$ \\
\hline
5.& \multicolumn{2}{l}{$(P^0 P^+ \bigcup V^0 P^+)$}   & 5.04/7 & $\left(81^{+36}_{-18}\right)^{\circ}$\\
6.& \multicolumn{2}{l}{$(P^0 P^+ \bigcup V^0 P^+ \bigcup P^0 V^+)$} & 5.08/9 & $\left(81^{+36}_{-18}\right)^{\circ}$  \\
7.& \multicolumn{2}{l}{$(P^0 P^+ \bigcup V^0 P^+ \bigcup V^0 V^+)$} & 5.27/9 & $\left(81^{+36}_{-18}\right)^{\circ}$  \\
\hline
8.& \multicolumn{2}{l}{$(P^0 P^+ \bigcup V^0 P^+ \bigcup P^0 V^+ \bigcup V^0 V^+)$} & 5.80/11 & $\left(82^{+35}_{-19}\right)^{\circ}$  \\
\hline
\hline
\end{tabular}
\end{center}
\end{table}

This is confirmed in the middle part of the table. 
%DS The best U-spin fit is achieved when $V^0 P^+$ and $P^0 P^+$ data are combined (30 data points) and fitted with 23 parameters (two sets of six U-spin amplitudes and five strong phases, plus the weak phase $\gamma$). 
The joint $(P^0 P^+ \bigcup V^0 P^+)$ fit prefers the same value of $\gamma$ as the $P^0 P^+$ one, namely, $\gamma=(81^{+36}_{-18})^{\circ}$. The addition of other subsets does not change this result, as expected. Thus, at the moment the results of all joint fits, including the full $(P^0 P^+ \bigcup  V^0 P^+ \bigcup P^0 V^+ \bigcup V^0  V^+)$ fit, are predominantly determined by the current $P^0 P^+$ data and produce practically identical results.

The above results are based on the world averages for branching ratios and $CP$ asymmetries in charged charmless $B$ decays. When the individual values from BaBar and Belle are very different, we employed the PDG scaling factor $S$ to boost uncertainties on the weighted averages, as shown in Appendix A. This modification only slightly affects the final result. The joint U-spin $(P^0 P^+ \bigcup  V^0 P^+)$ fit to the {\it unscaled} data prefers the same central value for the weak phase: $\gamma=(82^{+33}_{-17})^{\circ}$.

%DS restore if necessary
%\subsection{Future outlook \label{sec:future}}

\section{NEUTRAL DECAYS \label{sec:neutral}}

Charmless hadronic decays of $B^0$ and $B_s$ mesons to two-meson final states that contain vector or pseudoscalar mesons comprise seven subsets: $P^- P^+$, $V^- V^+$, $P^- V^+$, $V^- P^+$, $P^0 P^0$, $V^0 V^0$, 
and $V^0 P^0$. 

Unlike the charged $B^+$ meson which is a U-spin singlet, the neutral $B^0$ and $B_s$ belong to a U-spin doublet:
\be
\left[ \begin{array}{c} |\half~~\half \rangle \\
|\half -\!\half \rangle \end{array}
\right] = \left[ \begin{array}{c} |d \bar b \rangle = |B^0\rangle\\ 
\!| s \bar b \rangle = |B_s\rangle
\end{array} \right]~~.
\ee

The $\Delta S = 0$ Hamiltonian transforms like a $\bar d \sim -|\half -\half \rangle$, while the  $|\Delta S| = 1$ effective Hamiltonian transforms like a $\bar s \sim |\half \half \rangle$. The effect of these Hamiltonians on the two neutral $B$ mesons is 
\bea
\begin{array}{llc}
\label{eq:Heff}
{\cal H}_{\rm eff}^{\bar b\to\bar d}\;|B^0\rangle & = & 
-\frac{1}{\sqrt2 } |1~~0\rangle+\frac{1}{\sqrt2 } |0~~0\rangle ~~,\\
{\cal H}_{\rm eff}^{\bar b\to\bar s}\;|B^0\rangle & = &  |1~~1\rangle ~~,\\
{\cal H}_{\rm eff}^{\bar b\to\bar d}\;|B_s\rangle & = &  -|1~~-1\rangle ~~,\\
{\cal H}_{\rm eff}^{\bar b\to\bar s}\;|B_s\rangle & = &  
\frac{1}{\sqrt2 } |1~~0\rangle+\frac{1}{\sqrt2 } |0~~0\rangle ~~.
\end{array}
\eea
Thus, unlike $B^+$ decays where the final state must have $U=\frac12$,
the final states of neutral $B$ decays have two options: they can be in both  $U=0$ and $U=1$ U-spin states.

\subsection{$B^0, B_s \to M^- M^+$  ($P^- P^+$, $V^- V^+$, $P^- V^+$, 
$V^- P^+$) decays \label{sec:neutral-to-two-charged} }

Charmless decays of $B^0$ and $B_s$ to two {\it charged} mesons belong to one of the four subsets: $P^- P^+$, $V^- V^+$, $P^- V^+$, and $V^- P^+$. Each of the subsets consists of six decays. For example, three of neutral $B \to P^- P^+$ decays are $B^0 \to \pi^- \pi^+, K^- K^+, \pi^- K^+$. The other three are the exact U-spin mirror images of these decays, that is, 
$B_s \to K^- K^+, \pi^- \pi^+,  K^- \pi^+$. 

%DS Discuss Bose symmetry??? U-spin breaking? Khodjamirian paper

Since both charged mesons of the final state belong to a U-spin doublet (Eq.~(\ref{eqn:doublet}) and~(\ref{eqn:doublet2})) the final state can be either $U=0$ or $U=1$ for $\pi\pi$ and $KK$ decays and only $U=1$ for $\pi K$ decays. One can calculate that the actual physical amplitudes are:

\bea
\begin{array}{lllllllccccc}
A_{B^0}(\pi^- \pi^+ &,& \rho^- \rho^+ &,& \pi^- \rho^+ &,& \rho^- \pi^+ &)\, = &-&\frac12 A_1^d &+& \frac12 A_0^d,\\ 
A_{B^0}(K^- K^+ &,& K^{*-} K^{*+} &,& K^- K^{*+} &,& K^{*-} K^+ &)\, = &+&
\frac12 A_1^d &+& \frac12 A_0^d,\\ 
A_{B^0}(\pi^- K^+ &,& \rho^- K^{*+} &,& \pi^- K^{*+} &,& \rho^- K^+ &)\, = &-&  A_1^s &&\quad\quad,\\
A_{B_s}(K^- K^+ &,& K^{*-} K^{*+} &,& K^- K^{*+} &,& K^{*-} K^+ &)\, = &-&\frac12 A_1^s &+& \frac12 A_0^s,\\ 
A_{B_s}(\pi^- \pi^+ &,& \rho^- \rho^+ &,& \pi^- \rho^+ &,& \rho^- \pi^+ &)\, = &+& \frac12 A_1^s &+& \frac12 A_0^s,\\ 
A_{B_s}(K^- \pi^+ &,& K^{*-} \rho^+ &,& K^- \rho^+ &,& K^{*-} \pi^+ &)\, = &-&  A_1^d &&\quad\quad,\\  
\end{array}
\eea
where subscripts $1$ and $0$ refer to $U_f$, the U-spin value of the final state.

Decay amplitudes of the pairs that are directly related to each other by U-spin symmetry ($A_{B^0}(\pi^- \pi^+)$ and $A_{B_s}(K^- K^+)$, for instance) are virtually identical with the only exception of subscripts $d$ and $s$ that identify strangeness-conserving and strangeness-changing transformations. When these amplitudes are expanded as sums of ``u-term" and ``c-term" amplitudes (Eq.~(\ref{eqn:Ad}) and~(\ref{eqn:As})) these decay amplitudes feature the same U-spin amplitudes and strong phases multiplied by different CKM parameters.

$A_1^d$ and $A_1^s$ are expressed in terms of the same $A_1^u$ ``u-term" amplitude and $A_1^c$ ``c-term" amplitude, the similar statement is also true for $A_0^d$ and $A_0^s$. So, the description of neutral $B \to M^- M^+$ decays involves four U-spin amplitudes in total. Three relative strong phases and the weak phase $\gamma$ complete the list of parameters. Thus, each of the four decay subsets is separately described by 8 parameters. For a meaningful fit to the experimental data, eight or more subset data points must be available.

\subsection{Experimental data \label{sec:neutral-to-two-charged exp data}} 
Below we summarize the experimental data on $M^- M^+$ decays that is currently available. We also estimate which data modes may potentially get measured in the near future based on QCD factorization approach predictions for $PP$ and $PV$ decay modes~\cite{Beneke:2003zv}. 
%DS WHAT ABOUT VV PREDICTIONS??? 
We assume that a branching ratio will soon be measured if the preferred S4 scenario in~\cite{Beneke:2003zv} predicts that it is larger than $0.5\cdot10^{-6}$. We also assume that the  direct $CP$ asymmetry of a neutral $B$ decay will be measured if its branching ratio is larger than $1\cdot10^{-6}$. The mixing-induced $CP$ asymmetry requires time-dependent measurements so we assume that these will only be measured for the decays whose branching ratio is larger than $2\cdot10^{-6}$.

\begin{enumerate}

\item  
$P^- P^+$ decays: 9 data points (currently available). Potentially available (based on QCD factorization predictions, as explained above): 12 data points. 
\begin{itemize}
\item
$B^0$ decays: 6 data points \quad (2 branching ratios, 1 mixing induced and 2 direct $CP$ asymmetries, 1 upper limit). Potentially: no new measurements are expected but the experimental accuracy will improve.
\item
%DS ask Soni: will Tevatron be able to do time-dependent measurements? 
$B_s$ decays: 3 data points \quad (1 branching ratio, 2 upper limits). Potentially: 6 data points \quad (2 branching ratios, 1 mixing induced and 2 direct $CP$ asymmetries, 1 upper limit).
\end{itemize}

\item  
$V^- V^+$ decays: 5 data points. 
%DS (DS:Predictions for VV?)
\begin{itemize}
\item
$B^0$ decays: 5 data points \quad (1 branching ratio, 1 mixing induced and 1 direct $CP$ asymmetry, 2 upper limits).
\item
$B_s$ decays: none.
\end{itemize}

\item
$V^- P^+$ decays: 4 data points. Potentially: 10 data points.
\begin{itemize}
\item
$B^0$ decays: 4 data points \quad (2 branching ratios, 2 direct $CP$ asymmetries). Potentially: 5 data points \quad (2 branching ratios, 2 direct $CP$ asymmetries, 1 upper limit).
\item
$B_s$ decays: none. Potentially: 5 data points \quad (2 branching ratios, 2 direct $CP$ asymmetries, 1 upper limit).
\end{itemize}

\item
$P^- V^+$ decays: 4 data points. Potentially: 10 data points.
\begin{itemize}
\item
$B^0$ decays: 4 data points \quad (2 branching ratios, 2 direct $CP$ asymmetries). Potentially: 5 data points \quad (2 branching ratios, 2 direct $CP$ asymmetries, 1 upper limit).
\item
$B_s$ decays: none. Potentially: 5 data points \quad (2 branching ratios, 2 direct $CP$ asymmetries, 1 upper limit).
\end{itemize}

\item
$V^- P^+$ and $P^- V^+$ decays together: 10 data points. Potentially: 22 data points.
\begin{itemize}
\item
$B^0$ decays: 10 data points \quad (6 measurements in the 
$\rho^{\mp} \pi^{\pm}$ system, including $S$ and $\Delta S$, 4 measurements in $\rho^- K^+$ and $\pi^- K^{*+}$ decays). Potentially: 12 data points \quad (2 additional upper limits on $K^{*-} K^+$ and $K^- K^{*+}$).
\item
$B_s$ decays: none. Potentially: 10 data points \quad (4 branching ratios, 4 direct $CP$ asymmetries, 2 upper limits).
\end{itemize}

\end{enumerate}

Considering that one needs at least 8 parameters for each of the first four subsets and 15 parameters for the joint $(V^- P^+)\bigcup(P^- V^+)$ subset, neither of them has or will ever have enough data points in the $B^0$ subsector alone. $B_s$ data is crucial and must be used for a reasonable fit. Each of the four subsets will eventually have at least 10 data points and will allow the extraction of the CKM angle $\gamma$.

At the moment the $B^0, B_s \to P^- P^+$ subset is the only one with more than 8 available data points. A U-spin fit to this data finds two local minima and prefers values of $\gamma$ to lie at $\gamma=(37\pm3)^{\circ}$ or $\gamma=(80^{+6}_{-8})^{\circ}$. These two minima lie at about the same $\chi^2$ level of about 3.6/1, see Table~\ref{tab:results}.

\begin{table}[tp]
\begin{center}
\small
\caption{Results of the U-spin fits to charged and neutral subsets of charmless $B \to M_1 M_2$ decays. The bottom panel shows $\gamma$ as determined from direct measurements in $B \to D^{(*)} K^{(*)}$ decays, from indirect constraints on the apex of the unitarity triangle, and from SU(3) fits to charmless $PP$ and $VP$ decays, for comparison purposes.
\label{tab:results}}
\begin{tabular}{clccc}
\hline
\hline
%\quad Fit~\quad\quad & Subset & Modes &\quad $\chi^2/dof$~\quad\quad & \quad\quad $\gamma$~\quad\quad\quad \\
Fit & Subset  & $\chi^2/dof$ & $\gamma$\\
\hline
1.& $B^+ \to P^0 P^+$ & 3.2/4 & $\left(81^{+36}_{-18}\right)^{\circ}$  \\
2.& $B^0, B_s \to P^- P^+$ (two minima): & 3.6/1 & $\left(80^{+6}_{-8}\right)^{\circ}$   \\
& & 3.7/1 &  $\left(37\pm3\right)^{\circ}$ \\
\hline
3. & $(B^+ \to P^0 P^+) \bigcup (B^0, B_s \to P^- P^+)$~\quad\quad  & 6.8/6 & $\left(80^{+6}_{-8}\right)^{\circ}$  \\
\hline
\hline
\multicolumn{3}{l}{Direct measurements, BaBar~\cite{Aubert:2005yj}}
 & $(67\pm28\pm13\pm11)^{\circ}$ \\
\multicolumn{3}{l}{Direct measurements, Belle~\cite{Poluektov:2006ia}} 
 & $\left(53^{+15}_{-18}\pm3\pm9\right)^{\circ}$ \\
\multicolumn{3}{l}{Indirect constraints, CKMFitter~\cite{Charles:2004jd}} &  $\left(59.8^{+4.9}_{-4.2}\right)^{\circ}$\\
\multicolumn{3}{l}{Indirect constraints, UTFit~\cite{Bona:2006ah}} &  $\left(61.3\pm4.5\right)^{\circ}$\\
\multicolumn{3}{l}{SU(3) fits to $VP$ decays~\cite{Suprun:2006}} & 
$\left(66.2^{+3.8}_{-3.9}\pm0.1\right)^{\circ}$\\
\multicolumn{3}{l}{SU(3) fits to $PP$ decays~\cite{Suprun:2006}} &  $\left(59\pm9\pm2\right)^{\circ}$\\
\hline
\hline
\end{tabular}
\end{center}
\end{table}

To resolve this ambiguity we combine this U-spin fit with the $B^+ \to P^0 P^+$ fit, the one that dominates all joint $B^+$ fits. That fit has its deepest minimum at $\gamma=81^\circ$. Naturally, the ambiguity of the 
$B^0, B_s \to P^- P^+$ fit results gets resolved in the joint $(B^+ \to P^0 P^+ \bigcup B^0, B_s \to P^- P^+)$ fit in favor of large $\gamma$. This joint fit has two minima: $\gamma=(38\pm3)^{\circ}$, $\chi^2=15.1/6$ and the much deeper one at $\gamma=(80^{+6}_{-8})^{\circ}$, $\chi^2=6.8/6$. The latter solution is the main result of the current U-spin fits in this paper. 

\subsection{$B^0, B_s \to M^0 M^0$  ($V^0 P^0$, $P^0 P^0$, $V^0 V^0$) decays. \label{sec:neutral-to-neutral}}

%DS MENTION: 3 subsets ($B^0, B_s \to V^0 P^0$, $B^0, B_s \to P^0 P^0$, $B^0, B_s \to V^0 V^0$) with the same meaning for U-spin amplitudes within each subset.

In Section~\ref{sec:uspin} charmless neutral $M^0$ mesons (both pseudoscalar and vector ones) were shown to consist of three U-spin components: the  $|1~~0\rangle$ triplet state, the U-spin singlet $|0~~0\rangle_8$ state and the SU(3) singlet $|0~~0\rangle_1$ state. In charged decays each of these three components may combine with the doublet of the charged meson and thus, three amplitudes, $A^d_1$, $A^d_{0_8}$, $A^d_{0_1}$, comprise all possible contributions to the $\Delta S = 0$ decay amplitudes. Each of these three amplitudes has a ``tree" and a ``penguin" component: $A^u_1$ and $A^c_1$, $A^u_{0_8}$ and $A^c_{0_8}$, $A^u_{0_1}$ and $A^c_{0_1}$. Three other  amplitudes, $A^s_1$, $A^s_{0_8}$, $A^s_{0_1}$, that describe charged  $|\Delta S| = 1$ decays, consist of the same six components: the same three ``trees" and the same three ``penguins". 

The U-spin description of the $B^0, B_s \to M^0 M^0$ decays is more complicated because the final state consists of two neutral mesons. Each of them may have as many as three multiplet components. Four neutral $K$ mesons ($K^0$, $K^{*0}$, and their antiparticles) are pure triplets in terms of U-spin, either $|1~~1\rangle$ or $|1~~-1\rangle$. $\pi^0$ and $\rho^0$ mesons receive contributions from the $|1~~0\rangle$ triplet state and from the U-spin singlet $|0~~0\rangle_8$ state. The other four mesons, $\eta$, $\eta'$, $\omega$, and $\phi$, contain all three multiplet states: $|1~~0\rangle$, $|0~~0\rangle_8$, and $|0~~0\rangle_1$. When both final state $M^0$ mesons are neutral vector or pseudoscalar $K$ mesons (one with the $|1~~1\rangle$ U-spin, the other with $|1~~-1\rangle$) then U-spin of the final state can only be either $|1~~0\rangle$ or $|0~~0\rangle$. The $|2~~0\rangle$ state cannot couple to 
${\cal H}_{\rm eff}^{\bar b\to\bar d}\;|B^0\rangle$ or 
${\cal H}_{\rm eff}^{\bar b\to\bar s}\;|B_s\rangle$, as one can see from Eq.~(\ref{eq:Heff}). 

We introduce a general notation for all U-spin decay amplitudes. Each amplitude will have a superscript $d$ or $s$, depending on whether the decay is strangeness-conserving ($\bar b \to \bar d$) or strangeness-changing 
($\bar b \to \bar s$). The amplitudes will also feature two or three subscripts. The first two will show which U-spin multiplet each of the two final state mesons belong to. With most neutral mesons being linear combinations of several U-spin multiplet states, $B$ decays into these mesons will be described by several different U-spin amplitudes. We will use $A_{1,0_8}^{d,s}$ notation for U-spin amplitudes for decays into a triplet and an octet singlet, $A_{1,0_1}^{d,s}$ for triplet-SU(3) singlet amplitudes, $A_{0_8,0_8}^{d,s}$ for octet singlet-octet singlet, $A_{0_8,0_1}^{d,s}$ for octet singlet-SU(3) singlet, and $A_{0_1,0_1}^{d,s}$ for SU(3) singlet-SU(3) singlet amplitudes. Finally, as shown in the previous paragraph, triplet-triplet final states may have either U-spin  $U=1$ or $U=0$. They will be denoted by $A_{1,1,1}^{d,s}$ and $A_{1,1,0}^{d,s}$. 

Thus, the total number of different U-spin amplitudes that are involved in $M^0 M^0$ decays is 7. Each of these seven amplitudes also has a ``tree" and a ``penguin" component. So, 14 U-spin amplitudes are needed to describe all $B^0, B_s \to M^0 M^0$ decays. The total number of parameters that are required for the full fit to each of the three separate $M^0 M^0$ subsets ($V^0 P^0$, $P^0 P^0$ and $V^0 V^0$) is 28 (14 amplitudes, 13 relative strong phases and the weak phase $\gamma$). A subset must contain at least 28 experimental data points for a meaningful fit to be feasible.

\subsection{$B^0, B_s \to M^0 M^0$ decay amplitudes \label{sec:neutral-to-neutral amplitudes}}

%DS SHOW: 1 example of decay amplitude calculation? 

The full list of $B^0$ and $B_s$ decay amplitudes in terms of different U-spin components is shown below. In these equations each U-spin amplitude, for example $A_{1,1,1}$, has the same meaning within each subset ($B^0, B_s \to V^0 P^0$, for instance) but different meanings and values in each of the three subsets.   

\pagebreak

$B^0 \to M^0 M^0$ $\Delta S=0$ decays where $M$ is a $K$ meson (two contributing amplitudes: $A_{1,1,1}^d$ and $A_{1,1,0}^d$): 
\bea
\begin{array}{lccccccccc}
A_{B^0}(K^{*0} \overline{K^0}&,&  K^0 \overline{K^0}&,& &)\, = & 
+&\frac{A_{1,1,1}^d}{2}  &-& \frac{A_{1,1,0}^d}{\sqrt{6}} \\ A_{B^0}(\overline{K^{*0}} K^0&,& &,& \overline{K^{*0}} K^{*0} &)\, = & 
-&\frac{A_{1,1,1}^d}{2}  &-& \frac{A_{1,1,0}^d}{\sqrt{6}} \\ 
\end{array}
\eea

$B^0 \to M^0 M^0$ $|\Delta S|=1$ decays (three contributing amplitudes: $A_{1,1,1}^s$, $A_{1,0_8}^s$, $A_{1,0_1}^s$):  
\bea
\begin{array}{lccccccccccc}
A_{B^0}(K^{*0}\pi^0 &,&K^0\pi^0 &,& &)\, = &-&\frac{A_{1,1,1}^s}{2\sqrt{2}} &+&  \frac{3A_{1,0_8}^s}{2\sqrt{3}}~~,\\
A_{B^0}(K^{*0}\eta &,&K^0\eta &,& &)\, = &+& \frac{A_{1,1,1}^s}{\sqrt{3}} &+&  \frac{2A_{1,0_8}^s}{3\sqrt{2}} &-& \frac{A_{1,0_1}^s}{3}~~,\\
A_{B^0}(K^{*0}\eta' &,&K^0\eta' &,& &)\, =&+&  \frac{A_{1,1,1}^s}{2\sqrt{6}} &+&  \frac{A_{1,0_8}^s}{6} &+&  \frac{4A_{1,0_1}^s}{3\sqrt{2}}~~,\\
A_{B^0}(\rho^0 K^0&,& &,&\rho^0 K^{*0}&)\, = &+& \frac{A_{1,1,1}^s}{2\sqrt{2}} &+&  \frac{3A_{1,0_8}^s}{2\sqrt{3}}~~,\\
A_{B^0}(\omega K^0&,& &,&\omega K^{*0}&)\, = &+& \frac{A_{1,1,1}^s}{2\sqrt{2}} &-& \frac{A_{1,0_8}^s}{2\sqrt{3}} &+&  \frac{2A_{1,0_1}^s}{\sqrt{6}}~~,\\
A_{B^0}(\phi K^0&,& &,&\phi K^{*0}&)\, = &-& \frac{A_{1,1,1}^s}{2} &+&  \frac{A_{1,0_8}^s}{\sqrt{6}} &+&  \frac{A_{1,0_1}^s}{\sqrt3}~~.\\
\end{array}
\eea

$B^0 \to M^0 M^0$ $\Delta S=0$ decays where $M$ is not a $K$ meson: six contributing amplitudes; 
%DS
% $A_{1,1,0}^d$, $A_{1,0_8}^d$, $A_{1,0_1}^d$, $A_{0_8,0_8}^d$, $A_{0_8,0_1}^d$, $A_{0_1,0_1}^d$); 
$\sqrt{2}$ factor modifies amplitudes of decays into identical particles:
\bea
\begin{array}{lllllccccccccccccc}
A_{B^0}(\rho^0 \pi^0&,&&,&  &)\, = & 
-&\frac{A_{1,1,0}^d}{4\sqrt{6}}  &+& \frac{3 A_{1,0_8}^d}{2\sqrt{6}} && &+& \frac{3A_{0_8,0_8}^d}{4\sqrt{2}} & & & & \\ 
\sqrt{2}A_{B^0}(&,& \pi^0 \pi^0&,& \rho^0 \rho^0 &)\, = & 
-&\frac{A_{1,1,0}^d}{4\sqrt{6}}  &+& \frac{3 A_{1,0_8}^d}{2\sqrt{6}} && &+& \frac{3A_{0_8,0_8}^d}{4\sqrt{2}} & & & & \\ 
A_{B^0}(\rho^0 \eta&,& \pi^0 \eta&, &  &)\, = & 
+&\frac{A_{1,1,0}^d}{6} &-& \frac{A_{1,0_8}^d}{3} &-& 
\frac{A_{1,0_1}^d}{6\sqrt{2}}  &+& \frac{A_{0_8,0_8}^d}{2\sqrt{3}}  &-& 
\frac{A_{0_8,0_1}^d}{2\sqrt{6}} && \\ 
A_{B^0}(\rho^0 \eta'&,& \pi^0 \eta'&,& &)\, = & 
+&\frac{A_{1,1,0}^d}{12\sqrt{2}} &-& \frac{A_{1,0_8}^d}{6\sqrt{2}} &+& 
\frac{A_{1,0_1}^d}{3}  &+& \frac{A_{0_8,0_8}^d}{4\sqrt{6}}  &+& 
\frac{A_{0_8,0_1}^d}{\sqrt{3}} && \\ 
A_{B^0}(\omega \pi^0&, & &,& \omega \rho^0 &)\, = & 
-&\frac{A_{1,1,0}^d}{4\sqrt{6}} &+& \frac{A_{1,0_8}^d}{2\sqrt{6}} &+& 
\frac{A_{1,0_1}^d}{2\sqrt{3}}  &-& \frac{A_{0_8,0_8}^d}{4\sqrt{2}}  &+& 
\frac{A_{0_8,0_1}^d}{2}  &&  \\ 
A_{B^0}(\omega \eta&, & &, & &)\, = & 
+&\frac{A_{1,1,0}^d}{6} &+& \frac{A_{1,0_8}^d}{3} &-& 
\frac{5A_{1,0_1}^d}{6\sqrt{2}}  &-& \frac{A_{0_8,0_8}^d}{6\sqrt{3}}  &+& 
\frac{5A_{0_8,0_1}^d}{6\sqrt{6}}  &-& \frac{A_{0_1,0_1}^d}{3\sqrt{3}} \\ 
A_{B^0}(\omega \eta'&, & &, & &)\, = & 
+&\frac{A_{1,1,0}^d}{12\sqrt{2}} &+& \frac{A_{1,0_8}^d}{6\sqrt{2}} &+& 
\frac{A_{1,0_1}^d}{6}  &-& \frac{A_{0_8,0_8}^d}{12\sqrt{6}}  &-& 
\frac{A_{0_8,0_1}^d}{6\sqrt{3}}  &+& \frac{4A_{0_1,0_1}^d}{3\sqrt{6}} \\ 
A_{B^0}(\phi \pi^0&, & &, &\phi \rho^0 &)\, = & 
+&\frac{A_{1,1,0}^d}{4\sqrt{3}} &-& \frac{A_{1,0_8}^d}{2\sqrt{3}} &+& 
\frac{A_{1,0_1}^d}{2\sqrt{6}}  &+& \frac{A_{0_8,0_8}^d}{4}  &+& 
\frac{A_{0_8,0_1}^d}{2\sqrt{2}}  &&  \\ 
A_{B^0}(\phi \eta&,  & &, & &)\, = & 
-&\frac{A_{1,1,0}^d}{3\sqrt{2}} &-& \frac{2A_{1,0_8}^d}{3\sqrt{2}} &-& 
\frac{A_{1,0_1}^d}{6}  &+& \frac{A_{0_8,0_8}^d}{3\sqrt{6}}  &+& 
\frac{A_{0_8,0_1}^d}{6\sqrt{3}}  &-& \frac{A_{0_1,0_1}^d}{3\sqrt{6}} \\ 
A_{B^0}(\phi \eta'&, & &, & &)\, = & 
-&\frac{A_{1,1,0}^d}{12} &-& \frac{A_{1,0_8}^d}{6} &-& 
\frac{5A_{1,0_1}^d}{6\sqrt{2}}  &+& \frac{A_{0_8,0_8}^d}{12\sqrt{3}}  &+& 
\frac{5A_{0_8,0_1}^d}{6\sqrt{6}}  &+& \frac{2A_{0_1,0_1}^d}{3\sqrt{3}} \\ 
\sqrt{2}A_{B^0}( &, &\eta \eta&,& &)\, = & 
-&\frac{2A_{1,1,0}^d}{3\sqrt{6}} &-& \frac{4A_{1,0_8}^d}{3\sqrt{6}} &+& 
\frac{2A_{1,0_1}^d}{3\sqrt{3}}  &+& \frac{2A_{0_8,0_8}^d}{9\sqrt{2}}  &-& 
\frac{2A_{0_8,0_1}^d}{9}  &+& \frac{A_{0_1,0_1}^d}{9\sqrt{2}} \\ 
A_{B^0}( &, &\eta \eta'&, & &)\, = & 
-&\frac{A_{1,1,0}^d}{6\sqrt{3}} &-& \frac{A_{1,0_8}^d}{3\sqrt{3}} &-& 
\frac{7A_{1,0_1}^d}{6\sqrt{6}}  &+& \frac{A_{0_8,0_8}^d}{18}  &+& 
\frac{7A_{0_8,0_1}^d}{18\sqrt{2}}  &-& \frac{2A_{0_1,0_1}^d}{9} \\ 
\sqrt{2}A_{B^0}( &, &\eta' \eta'&, & &)\, = & 
-&\frac{A_{1,1,0}^d}{12\sqrt{6}} &-& \frac{A_{1,0_8}^d}{6\sqrt{6}} &-& 
\frac{2A_{1,0_1}^d}{3\sqrt{3}}  &+& \frac{A_{0_8,0_8}^d}{36\sqrt{2}}  &+& 
\frac{2A_{0_8,0_1}^d}{9}  &+& \frac{8A_{0_1,0_1}^d}{9\sqrt{2}} \\ 
\sqrt{2}A_{B^0}(&, & &,& \omega \omega &)\, = & 
-&\frac{A_{1,1,0}^d}{4\sqrt{6}} &-& \frac{A_{1,0_8}^d}{2\sqrt{6}} &+& 
\frac{A_{1,0_1}^d}{\sqrt{3}}  &+& \frac{A_{0_8,0_8}^d}{12\sqrt{2}}  &-& 
\frac{A_{0_8,0_1}^d}{3}  &+& \frac{2A_{0_1,0_1}^d}{3\sqrt{2}}  \\ 
A_{B^0}(&, & &,& \omega \phi &)\, = & 
+&\frac{A_{1,1,0}^d}{4\sqrt{3}} &+& \frac{A_{1,0_8}^d}{2\sqrt{3}} &-& 
\frac{A_{1,0_1}^d}{2\sqrt{6}}  &-& \frac{A_{0_8,0_8}^d}{12}  &+& 
\frac{A_{0_8,0_1}^d}{6\sqrt{2}}  &+& \frac{A_{0_1,0_1}^d}{3}  \\ 
\sqrt{2}A_{B^0}(&, & &,& \phi \phi &)\, = & 
-&\frac{A_{1,1,0}^d}{2\sqrt{6}} &-& \frac{A_{1,0_8}^d}{\sqrt{6}} &-& 
\frac{A_{1,0_1}^d}{\sqrt{3}}  &+& \frac{A_{0_8,0_8}^d}{6\sqrt{2}}  &+& 
\frac{A_{0_8,0_1}^d}{3}  &+& \frac{2A_{0_1,0_1}^d}{3\sqrt{2}}  \\ 
\end{array}
\eea

\pagebreak

$B_s \to M^0 M^0$ $|\Delta S|=1$ decays where $M$ is a $K$ meson (two contributing amplitudes: $A_{1,1,1}^s$ and $A_{1,1,0}^s$): 
\bea
\begin{array}{lccccccccc}
A_{B_s}(\overline{K^{*0}} K^0&,& \overline{K^0} K^0&,&  &)\, = & 
+&\frac{A_{1,1,1}^s}{2}  &-& \frac{A_{1,1,0}^s}{\sqrt{6}} \\ 
A_{B_s}(K^{*0} \overline{K^0}&,&  &,& K^{*0} \overline{K^{*0}}  &)\, = & 
-&\frac{A_{1,1,1}^s}{2}  &-& \frac{A_{1,1,0}^s}{\sqrt{6}} \\ 
\end{array}
\eea

$B_s \to M^0 M^0$ $\Delta S=0$ decays (three contributing amplitudes: $A_{1,1,1}^d$, $A_{1,0_8}^d$, $A_{1,0_1}^d$):  
\bea
\begin{array}{lccccccccccc}
A_{B_s}(\overline{K^{*0}} \pi^0 &,&\overline{K^0} \pi^0 &,& &)\, = &+&\frac{A_{1,1,1}^d}{2\sqrt{2}} &+&  \frac{3A_{1,0_8}^d}{2\sqrt{3}}~~,\\
A_{B_s}(\overline{K^{*0}} \eta &,&\overline{K^0} \eta &,& &)\, = &-& \frac{A_{1,1,1}^d}{\sqrt{3}} &+&  \frac{2A_{1,0_8}^d}{3\sqrt{2}} &-& \frac{A_{1,0_1}^d}{3}~~,\\
A_{B_s}(\overline{K^{*0}} \eta' &,&\overline{K^0} \eta' &,& &)\, =&-&  \frac{A_{1,1,1}^d}{2\sqrt{6}} &+&  \frac{A_{1,0_8}^d}{6} &+&  \frac{4A_{1,0_1}^d}{3\sqrt{2}}~~,\\
A_{B_s}(\rho^0 \overline{K^0} &,& &,&\rho^0 \overline{K^{*0}} &)\, = &-& \frac{A_{1,1,1}^d}{2\sqrt{2}} &+&  \frac{3A_{1,0_8}^d}{2\sqrt{3}}~~,\\
A_{B_s}(\omega \overline{K^0} &,& &,&\omega \overline{K^{*0}} &)\, = &-& \frac{A_{1,1,1}^d}{2\sqrt{2}} &-& \frac{A_{1,0_8}^d}{2\sqrt{3}} &+&  \frac{2A_{1,0_1}^d}{\sqrt{6}}~~,\\
A_{B_s}(\phi \overline{K^0} &,& &,&\phi \overline{K^{*0}} &)\, = &+& \frac{A_{1,1,1}^d}{2} &+&  \frac{A_{1,0_8}^d}{\sqrt{6}} &+&  \frac{A_{1,0_1}^d}{\sqrt3}~~.\\
\end{array}
\eea

$B_s \to M^0 M^0$ $|\Delta S|=1$ decays where $M$ is not a $K$ meson: six contributing amplitudes;
%DS
% $A_{1,1,0}^s$, $A_{1,0_8}^s$, $A_{1,0_1}^s$, $A_{0_8,0_8}^s$, $A_{0_8,0_1}^s$, $A_{0_1,0_1}^s$): 
$\sqrt{2}$ factor modifies amplitudes of decays into identical particles:
\bea
\begin{array}{lllllccccccccccccc}
A_{B_s}(\rho^0 \pi^0&,& &,&  &)\, = & 
-&\frac{A_{1,1,0}^s}{4\sqrt{6}}  &-& \frac{3 A_{1,0_8}^s}{2\sqrt{6}} && &+& \frac{3A_{0_8,0_8}^s}{4\sqrt{2}} & & & & \\ 
\sqrt{2}A_{B_s}(&,& \pi^0 \pi^0&,& \rho^0 \rho^0 &)\, = & 
-&\frac{A_{1,1,0}^s}{4\sqrt{6}}  &-& \frac{3 A_{1,0_8}^s}{2\sqrt{6}} && &+& \frac{3A_{0_8,0_8}^s}{4\sqrt{2}} & & & & \\ 
A_{B_s}(\rho^0 \eta&,& \pi^0 \eta&, &  &)\, = & 
+&\frac{A_{1,1,0}^s}{6} &+& \frac{A_{1,0_8}^s}{3} &+& 
\frac{A_{1,0_1}^s}{6\sqrt{2}}  &+& \frac{A_{0_8,0_8}^s}{2\sqrt{3}}  &-& 
\frac{A_{0_8,0_1}^s}{2\sqrt{6}} && \\ 
A_{B_s}(\rho^0 \eta'&,& \pi^0 \eta'&,& &)\, = & 
+&\frac{A_{1,1,0}^s}{12\sqrt{2}} &+& \frac{A_{1,0_8}^s}{6\sqrt{2}} &-& 
\frac{A_{1,0_1}^s}{3}  &+& \frac{A_{0_8,0_8}^s}{4\sqrt{6}}  &+& 
\frac{A_{0_8,0_1}^s}{\sqrt{3}} && \\ 
A_{B_s}(\omega \pi^0&, & &,& \omega \rho^0 &)\, = & 
-&\frac{A_{1,1,0}^s}{4\sqrt{6}} &-& \frac{A_{1,0_8}^s}{2\sqrt{6}} &-& 
\frac{A_{1,0_1}^s}{2\sqrt{3}}  &-& \frac{A_{0_8,0_8}^s}{4\sqrt{2}}  &+& 
\frac{A_{0_8,0_1}^s}{2}  &&  \\ 
A_{B_s}(\omega \eta&, & &, & &)\, = & 
+&\frac{A_{1,1,0}^s}{6} &-& \frac{A_{1,0_8}^s}{3} &+& 
\frac{5A_{1,0_1}^s}{6\sqrt{2}}  &-& \frac{A_{0_8,0_8}^s}{6\sqrt{3}}  &+& 
\frac{5A_{0_8,0_1}^s}{6\sqrt{6}}  &-& \frac{A_{0_1,0_1}^s}{3\sqrt{3}} \\ 
A_{B_s}(\omega \eta'&, & &, & &)\, = & 
+&\frac{A_{1,1,0}^s}{12\sqrt{2}} &-& \frac{A_{1,0_8}^s}{6\sqrt{2}} &-& 
\frac{A_{1,0_1}^s}{6}  &-& \frac{A_{0_8,0_8}^s}{12\sqrt{6}}  &-& 
\frac{A_{0_8,0_1}^s}{6\sqrt{3}}  &+& \frac{4A_{0_1,0_1}^s}{3\sqrt{6}} \\ 
A_{B_s}(\phi \pi^0&, & &, &\phi \rho^0 &)\, = & 
+&\frac{A_{1,1,0}^s}{4\sqrt{3}} &+& \frac{A_{1,0_8}^s}{2\sqrt{3}} &-& 
\frac{A_{1,0_1}^s}{2\sqrt{6}}  &+& \frac{A_{0_8,0_8}^s}{4}  &+& 
\frac{A_{0_8,0_1}^s}{2\sqrt{2}}  &&  \\ 
A_{B_s}(\phi \eta&,  & &, & &)\, = & 
-&\frac{A_{1,1,0}^s}{3\sqrt{2}} &+& \frac{2A_{1,0_8}^s}{3\sqrt{2}} &+& 
\frac{A_{1,0_1}^s}{6}  &+& \frac{A_{0_8,0_8}^s}{3\sqrt{6}}  &+& 
\frac{A_{0_8,0_1}^s}{6\sqrt{3}}  &-& \frac{A_{0_1,0_1}^s}{3\sqrt{6}} \\ 
A_{B_s}(\phi \eta'&, & &, & &)\, = & 
-&\frac{A_{1,1,0}^s}{12} &+& \frac{A_{1,0_8}^s}{6} &+& 
\frac{5A_{1,0_1}^s}{6\sqrt{2}}  &+& \frac{A_{0_8,0_8}^s}{12\sqrt{3}}  &+& 
\frac{5A_{0_8,0_1}^s}{6\sqrt{6}}  &+& \frac{2A_{0_1,0_1}^s}{3\sqrt{3}} \\ 
\sqrt{2}A_{B_s}( &, &\eta \eta&,& &)\, = & 
-&\frac{2A_{1,1,0}^s}{3\sqrt{6}} &+& \frac{4A_{1,0_8}^s}{3\sqrt{6}} &-& 
\frac{2A_{1,0_1}^s}{3\sqrt{3}}  &+& \frac{2A_{0_8,0_8}^s}{9\sqrt{2}}  &-& 
\frac{2A_{0_8,0_1}^s}{9}  &+& \frac{A_{0_1,0_1}^s}{9\sqrt{2}} \\ 
A_{B_s}( &, &\eta \eta'&, & &)\, = & 
-&\frac{A_{1,1,0}^s}{6\sqrt{3}} &+& \frac{A_{1,0_8}^s}{3\sqrt{3}} &+& 
\frac{7A_{1,0_1}^s}{6\sqrt{6}}  &+& \frac{A_{0_8,0_8}^s}{18}  &+& 
\frac{7A_{0_8,0_1}^s}{18\sqrt{2}}  &-& \frac{2A_{0_1,0_1}^s}{9} \\ 
\sqrt{2}A_{B_s}( &, &\eta' \eta'&, & &)\, = & 
-&\frac{A_{1,1,0}^s}{12\sqrt{6}} &+& \frac{A_{1,0_8}^s}{6\sqrt{6}} &+& 
\frac{2A_{1,0_1}^s}{3\sqrt{3}}  &+& \frac{A_{0_8,0_8}^s}{36\sqrt{2}}  &+& 
\frac{2A_{0_8,0_1}^s}{9}  &+& \frac{8A_{0_1,0_1}^s}{9\sqrt{2}} \\ 
\sqrt{2}A_{B_s}(&, & &,& \omega \omega &)\, = & 
-&\frac{A_{1,1,0}^s}{4\sqrt{6}} &+& \frac{A_{1,0_8}^s}{2\sqrt{6}} &-& 
\frac{A_{1,0_1}^s}{\sqrt{3}}  &+& \frac{A_{0_8,0_8}^s}{12\sqrt{2}}  &-& 
\frac{A_{0_8,0_1}^s}{3}  &+& \frac{2A_{0_1,0_1}^s}{3\sqrt{2}}  \\ 
A_{B_s}(&, & &,& \omega \phi &)\, = & 
+&\frac{A_{1,1,0}^s}{4\sqrt{3}} &-& \frac{A_{1,0_8}^s}{2\sqrt{3}} &+& 
\frac{A_{1,0_1}^s}{2\sqrt{6}}  &-& \frac{A_{0_8,0_8}^s}{12}  &+& 
\frac{A_{0_8,0_1}^s}{6\sqrt{2}}  &+& \frac{A_{0_1,0_1}^s}{3}  \\ 
\sqrt{2}A_{B_s}(&, & &,& \phi \phi &)\, = & 
-&\frac{A_{1,1,0}^s}{2\sqrt{6}} &+& \frac{A_{1,0_8}^s}{\sqrt{6}} &+& 
\frac{A_{1,0_1}^s}{\sqrt{3}}  &+& \frac{A_{0_8,0_8}^s}{6\sqrt{2}}  &+& 
\frac{A_{0_8,0_1}^s}{3}  &+& \frac{2A_{0_1,0_1}^s}{3\sqrt{2}}  \\ 
\end{array}
\eea

While the number of parameters that are needed to describe each of the three big subsets ($V^0 P^0$, $P^0 P^0$, $V^0 V^0$) is 28, one may try to creatively choose smaller subsubsets with fewer amplitudes involved. One natural possibility can be three
$(B^0 \to M^0 M^0, |\Delta S|=1) \bigcup (B_s \to M^0 M^0, \Delta S=0)$ subsubsets with three contributing amplitudes: $A_{1,1,1}$, $A_{1,0_8}$, $A_{1,0_1}$. Each of these features ``tree" and ``penguin" pieces that are multiplied by different CKM factors to form either $A^d$ or $A^s$ version of the above three amplitudes. With 6 amplitudes, 5 relative strong phases and the CKM angle $\gamma$, one needs 12 parameters for the full description of this subsubset. Depending on which data points are available (or will soon become available), exploring a smaller subsubset may have some practical advantages.

\subsection{Experimental data. \label{sec:neutral-to-neutral exp data}} 

%DS UPDATE using new HFAG Winter 2006 data.
Below we summarize the experimental data on $M^0 M^0$ decays that is currently available. 
We also estimate which data modes may potentially get measured in the near future based on QCD factorization predictions, just as we did in Chapter~\ref{sec:neutral-to-two-charged exp data}. For $B_s$ decays the criteria for potential measurement of branching ratios and asymmetries are set twice as high as the corresponding thresholds for $B^0$ decays.

\begin{enumerate}

\item
$B^0 \to V^0 P^0$ decays: 27 data points (currently available). Potentially available (based on QCD factorization predictions, as explained in Chapter~\ref{sec:neutral-to-two-charged exp data}): 29 data points.
\begin{itemize}
\item
$\Delta S = 0$: 12 data points \quad (1 branching ratio, 1 direct $CP$ asymmetry, 10 upper limits). Potentially: 12 data points \quad (4 branching ratio, 1 direct $CP$ asymmetry, 7 upper limits).
\item
$|\Delta S| = 1$: 15 data points \quad (4 branching ratios, 3 mixing induced and 6 direct $CP$ asymmetries, 2 upper limits). Potentially: 17 data points \quad (6 branching ratios, 5 mixing induced and 6 direct $CP$ asymmetries).
\end{itemize}

\item  
$B^0 \to P^0 P^0$ decays: 17 data points. Potentially: 19 data points. 

\begin{itemize}
\item
$\Delta S = 0$: 10 data points \quad (2 branching ratios, 1 mixing induced and 2 direct $CP$ asymmetries, 5 upper limits). Potentially: 11 data points \quad (2 branching ratios, 2 mixing induced and 2 direct $CP$ asymmetries, 5 upper limits).
\item
$|\Delta S| = 1$: 7 data points \quad (3 branching ratios, 2 mixing induced and 2 direct $CP$ asymmetries). Potentially: 8 data points \quad (3 branching ratios, 2 mixing induced and 3 direct $CP$ asymmetries).
\end{itemize}

\item  
$B^0 \to V^0 V^0$ decays: 12 data points. Potentially: 12 or more data points.
%DS (DS:Predictions for VV?)
\begin{itemize}
\item
$\Delta S = 0$: 7 data points \quad (1 branching ratio and 6 upper limits). Potentially: 7 or more data points (new data points may not be measured).
\item
$|\Delta S| = 1$: 5 data points \quad (2 branching ratio, 2 direct $CP$ asymmetries, 1 upper limit). Potentially: 5 or more data points (new data points may not be measured).
\end{itemize}

\item  
$B_s \to V^0 P^0$ decays: no data points. Potentially: 21 data points \quad (2 branching ratios, 2 mixing induced and 2 direct $CP$ asymmetries, 15 upper limits). 
\begin{itemize}
\item
$\Delta S = 0$: no data points. Potentially: 6 data points \quad (6 upper limits).
\item
$|\Delta S| = 1$: no data points. Potentially: 15 data points \quad (2 branching ratios, 2 mixing induced and 2 direct $CP$ asymmetries, 9 upper limits).
\end{itemize}

\item  
$B_s \to P^0 P^0$ decays: no data points. Potentially: 19 data points \quad (5 branching ratios, 4 mixing induced and 5 direct $CP$ asymmetries, 5 upper limits). 
\begin{itemize}
\item
$\Delta S = 0$: no data points. Potentially: 4 data points \quad (1 branching ratio, 1 direct $CP$ asymmetry, 2 upper limits).
\item
$|\Delta S| = 1$: no data points. Potentially: 15 data points \quad (4 branching ratios, 4 mixing induced and 4 direct $CP$ asymmetries, 3 upper limits). 
\end{itemize}

\item  
$B_s \to V^0 V^0$ decays: 1 data point \quad ($\phi\phi$ branching ratio). Potentially: at least 10 data points \quad (1 branching ratio and 9 upper limits). 
\begin{itemize}
\item
$\Delta S = 0$: no data points. Potentially: 3 data points \quad (3 upper limits).
\item
$|\Delta S| = 1$: 1 data point \quad (1 branching ratio). Potentially: 7 data points \quad (1 branching ratio and 6 upper limits).
\end{itemize}

\end{enumerate}

Thus, in the near future, when at least upper limits for all $B_s$ decay modes are established, one would have 50 available data points in the 
$B^0, B_s \to V^0 P^0$ subset, 38 data points in the $B^0, B_s \to P^0 P^0$ subset and 22 data points in the $B^0, B_s \to V^0 V^0$ subset. When enough new $B_s$ decay data will get measured we can expect that full 28 parameter fits will be feasible in both $V^0 P^0$ and $P^0 P^0$ subsets. In principle, one would also be able to do a simultaneous fit to both subsets with angle $\gamma$ as the only common parameter.

%DS (DELETE/MODIFY this paragraph?)  However, in the near future several additional data points may get measured in the $V^0 P^0$ sector. First of all, the $\omega \eta$, $K^{*0} \pi^0$, and $K^{*0} \eta'$ decays may get observed, turning the current upper limits into actual measurements. One can also expect the first measurements for $K^0 \ol {K}^{*0}$ and $K^{*0} \ol {K}^0$ decays for which not even an upper limit is known at the moment. We also expect that direct $CP$ asymmetries will be measured for the first time for the $K^{*0} \eta'$ and $\rho^0 K^0$ decays and mixing-induced $CP$ asymmetries will be measured in the $K^{*0} \pi^0$, $K^{*0} \eta$ and $\rho^0 K^0$ decays. When this happens one will be able to perform a full U-spin fit to the $V^0 P^0$ data.)

One may try to choose a smaller subsubset of decays that is fully described by fewer than 28 parameters. As we mentioned earlier, any of the three (strangeness-changing $B^0$ / strangeness-conserving $B_s$) decay subsubsets  is described by 12 parameters. When $B_s$ data becomes available, we will have 12 or more data points in both  
$(B^0 \to V^0 P^0, |\Delta S|=1) \bigcup (B_s \to V^0 P^0, \Delta S=0)$  decay subsubset (potentially 23 data points) and 
$(B^0 \to P^0 P^0, |\Delta S|=1) \bigcup (B_s \to P^0 P^0, \Delta S=0)$ decay subsubset (potentially 12 data points). Fits to each of these two subsubsets will provide a simple way to extract $\gamma$.

One of the two sub-subsets, 
$(B^0 \to V^0 P^0, |\Delta S|=1) \bigcup (B_s \to V^0 P^0, \Delta S=0)$, already has 15 available data points in the $B^0$ decay sector and none in the $B_s$ sector. One might think that a U-spin fit within the $B^0$ sector may be performed and the CKM angle $\gamma$ be determined. We will now show why this is not the case.

\subsection{$B^0 \to V^0 P^0, |\Delta S| = 1$ sub-subset. \label{sec:neutral-to-neutral V0P0}}

As we have shown above, there are only three amplitudes that contribute to $B^0 \to M^0 M^0, |\Delta S| = 1$ decays: triplet-triplet amplitude $A_{1,1,1}^s$, triplet-octet singlet amplitude $A_{1,0_8}^s$ and triplet-SU(3) singlet amplitude $A_{1,0_1}^s$. Each of these amplitudes consists of a ``tree" and a ``penguin" component.  The final state of a $B^0$ strangeness-changing $|\Delta S|= 1$ decay was shown in Eq.~(\ref{eq:Heff}) to be a $|1~~1\rangle$ U-spin state.

Consider, for instance, $B^0 \to \rho^0 K^0$. The final state is a combination of 
$\rho^0 = \frac{\sqrt3}{2} |0~~0 \rangle_8 - \frac12 |1~~0 \rangle$ and 
$K^0 = |1~~1\rangle$. The first term in $\rho^0$ gives rise to a $A_{1,0_8}^s$ contribution to the physical decay amplitude while the second term is responsible for a $A_{1,1,1}^s$ contribution. Using the Clebsch-Gordan coefficients, we determine that 
$A(\rho^0 K^0)=\frac{1}{2\sqrt2}A_{1,1,1}^s + \frac{3}{2\sqrt3}A_{1,0_8}^s$. In the same way one can calculate U-spin decomposition for all physical decay amplitudes. For simplicity of the following expressions one can absorb the $\frac{1}{2\sqrt6}$ factor into the definition of $A^s_{1,1,1}$, the $\frac16$ factor into the definition of $A^s_{1,0_8}$ and the $\frac13$ factor into the definition of $A^s_{1,0_1}$. Then we find:
\bea
\begin{array}{ccc}
\label{eqn:deltaS1}
A(K^{*0}\pi^0) & = & -\sqrt3 A_{1,1,1}^s + 3\sqrt3 A_{1,0_8}^s~~,\\
A(K^{*0}\eta) & = &  2\sqrt2 A_{1,1,1}^s + 2\sqrt2 A_{1,0_8}^s - A_{1,0_1}^s~~,\\
A(K^{*0}\eta') & = &  A_{1,1,1}^s + A_{1,0_8}^s + 2\sqrt2 A_{1,0_1}^s~~,\\
A(\rho^0 K^0) & = & \sqrt3 A_{1,1,1}^s + 3\sqrt3 A_{1,0_8}^s~~,\\
A(\omega K^0) & = & \sqrt3 A_{1,1,1}^s - \sqrt3 A_{1,0_8}^s + \sqrt6 A_{1,0_1}^s~~,\\
A(\phi K^0) & = &  -\sqrt6 A_{1,1,1}^s + \sqrt6 A_{1,0_8}^s + \sqrt3 A_{1,0_1}^s~~.\\
\end{array}
\eea

The fact that only amplitudes with $s$ superscripts ($A_{1,0_8}^s$, $A_{1,1,1}^s$, $A_{1,0_1}^s$) contribute to this sub-subset of neutral decays turns out to be a crucial disadvantage, as we show below. To perform a fit to the branching ratios and $CP$ asymmetries for decays~(\ref{eqn:deltaS1}) one can use two different sets of 12 parameters. One of them (set A) is the standard set of 6 amplitudes ($|A^u_{1,0_8}|$, $|A^c_{1,0_8}|$, $|A^u_{1,1,1}|$, $|A^c_{1,1,1}|$, $|A^u_{1,0_1}|$, $|A^c_{1,0_1}|$), 5 relative strong phases and the weak phase $\gamma$. Another set of 12 parameters (set B) can be chosen to consist of the other 6 amplitudes (3 amplitudes with $s$ superscripts and their $CP$-conjugates): $|A^s_{1,0_8}|$, $|\ol{A}^s_{1,0_8}|$, $|A^s_{1,1,1}|$, $|\ol{A}^s_{1,1,1}|$, $|B^s_{1,0_1}|$, $|\ol{B}^s_{1,0_1}|$ and 6 effective phases associated with them. The important point is that these 12 parameters are not independent. 

In set A any one of the six strong phases can be set to zero, the other five strong phases are defined with respect to this phase. Suppose zero strong phase is chosen to be associated with amplitude $A^u_{1,0_8}$. Then $A^u_{1,0_8}=|A^u_{1,0_8}|$ and the amplitudes $A_{1,0_8}^s$ and its $CP$-conjugate $\ol{A}_{1,0_8}^s$ can be written as
$$
|A_{1,0_8}^s|e^{i\delta_{A_{1,0_8}^s}} \equiv A_{1,0_8}^s 
= V^*_{ub}V_{us}A^u_{1,0_8} + V^*_{cb}V_{cs}A^c_{1,0_8}
= V^*_{ub}V_{us}|A^u_{1,0_8}| + V^*_{cb}V_{cs}|A^c_{1,0_8}|e^{i\delta}~~\\
$$
\bea
\label{eq:dir}
\hspace{7cm} = |A'^u_{1,0_8}|e^{i\gamma} + |A'^c_{1,0_8}|e^{i\delta},
\eea
$$
|\ol{A}_{1,0_8}^s|e^{i\delta_{\ol{A}_{1,0_8}^s}} \equiv \ol{A}_{1,0_8}^s = V_{ub}V^*_{us}A^u_{1,0_8} + V_{cb}V^*_{cs}A^c_{1,0_8}
= V_{ub}V^*_{us}|A^u_{1,0_8}| + V_{cb}V^*_{cs}|A^c_{1,0_8}|e^{i\delta}~~\\
$$
\bea
\label{eq:cpconj}
\hspace{7cm} = |A'^u_{1,0_8}|e^{-i\gamma} + |A'^c_{1,0_8}|e^{i\delta},
\eea
where $|A'^u_{1,0_8}| \equiv |V^*_{ub}V_{us}||A^u_{1,0_8}|$ and $|A'^c_{1,0_8}| \equiv |V^*_{cb}V_{cs}||A^c_{1,0_8}|$. Note that the real parts of the above equations are equal:
\be
Re(|A_{1,0_8}^s|e^{i\delta_{A_{1,0_8}^s}}) = |A'^u_{1,0_8}|\cos(\gamma)+|A'^c_{1,0_8}|\cos(\delta)
= Re(|\ol{A}_{1,0_8}^s|e^{i\delta_{\ol{A}_{1,0_8}^s}}).
\ee
Thus, any of the four parameters that appear in the left-hand sides of  Eqs.~(\ref{eq:dir}) and (\ref{eq:cpconj}) can be written as a function of the other three. That is, there is one fewer degree of freedom, and the neutral $|\Delta S| = 1$ decay amplitudes contain only 11 unknown parameters. 

The fit to neutral $|\Delta S| = 1$ may prefer some specific values of parameters $|A_{1,0_8}^s|$, $|\ol{A}_{1,0_8}^s|$, and $\delta_{A_{1,0_8}^s}$. Unfortunately, for any values of these three parameters there is a continuous set of $\gamma$ values that satisfies Eqs.~(\ref{eq:dir}) and (\ref{eq:cpconj}). Thus, the fit to neutral $B^0, |\Delta S| = 1$ branching ratios is not sensitive to the weak phase $\gamma$ at all, regardless of the number of available data points. Of course, this conclusion remains valid for the $|\Delta S| = 1$ sub-subsets of $P^0 P^0$ and $V^0 V^0$ decays, too. Only fits to a full subset ($P^0 P^0$, $V^0 P^0$, or $V^0 V^0$) may be sensitive to the weak phase $\gamma$. In the near future only the $V^0 P^0$ subset is expected to provide more than 24 data points that are needed for the full U-spin fit to both $\Delta S = 0$ and $|\Delta S| = 1$ decay modes.

\subsection{$(B^0 \to V^0 P^0, |\Delta S|=1) \bigcup (B_s \to V^0 P^0, \Delta S=0)$ sub-subset. \label{sec:neutral-to-neutral full V0P0}}

The problem of the $(B^0 \to V^0 P^0, |\Delta S|=1)$ fit's insensitivity to $\gamma$ could be fixed, had at least one experimental measurement of the corresponding $(B_s \to V^0 P^0, \Delta S=0)$ branching ratios had been made. At present, none of these $B_s$ data points are available. 

As a test, we used QCD factorization-based predictions~\cite{Beneke:2003zv} for six $(B_s \to V^0 P^0, \Delta S=0)$ branching ratios. Theoretical uncertainties given in that paper are of about the same size as predicted central values. To simulate future experimental measurements of these decays  these uncertainties were cut in half. The resulting joint U-spin fit to both $B^0$ and $B_s$ decay modes becomes sensitive to the CKM phase $\gamma$ and prefers $\gamma=(39^{+32}_{-117})^{\circ}$. This exercise shows that in principle even decays into two neutral mesons may potentially be used in the future for $\gamma$ extraction. 

%DS\section{DISCUSSION}

\section{SUMMARY \label{sec:summary}}

Thus, with current statistics, the best U-spin fits allow the 
determination of $\gamma$ from charmless $B$ decays with a good accuracy.
In particular, neutral $B$ decay data is consistent with $\gamma=(80^{+6}_{-8})^{\circ}$, as determined from $B^0, B_s \to P^- P^+$ subset. This value is reasonably consistent with the current indirect determinations that expect $\gamma$ to lie between $52^{\circ}$ and $74^{\circ}$ \cite{Charles:2004jd,Bona:2006ah}. 
Note that the intrinsic theoretical uncertainty associated with 
possible U-spin breaking effects is expected to be rather small and that U-spin symmetry is the only assumption that is made 
in this approach~\cite{Gronau:2004tm}. 
Clearly, as data with higher statistics becomes available, the 
statistical uncertainties on $\gamma$ will become even smaller. 

At the moment the difference between the four values of $\gamma$ 
extracted from the four $B^{\pm}$ decay subsets ($P^0 P^{\pm}$, $P^0 V^{\pm}$, $V^0 P^{\pm}$, $V^0 V^{\pm}$) is not very meaningful due to 
large uncertainties (Table~\ref{tab:fits}). When all branching 
ratios and $CP$ asymmetries in charged $B$ decays are 
experimentally determined with high accuracy, U-spin approach 
should enable extraction of  $\gamma$ quite precisely from 
each of the four subsets of data. 
The resulting spread in $\gamma$ values should be small and could perhaps be used to indicate the systematic errors inherent in the method due to residual U-spin breaking effects. The crucial advantage of the method is that the extraction of $\gamma$ is completely model independent and entirely data driven. Note also that unlike the use of isospin for $\alpha$, electroweak penguins are not a problem in our approach.
Penguin contributions are entering in an important way in this U-spin approach for getting $\gamma$. That means that this method is
sensitive to new physics in the loops. In contrast, recall
that the standard $B \to DK$ methods~\cite{schune_ichep05} involve only tree $B$ decays. Comparison of $\gamma$ from these two methods is therefore important for uncovering new physics.  

%DS Thus, the best U-spin fits allow the extraction of $\gamma$ with a reasonable precision and the preferred value is $\gamma=(54^{+12}_{-11})^{\circ}$. Not only are the uncertainties quite small but also the central value $\gamma=54^{\circ}$ is very consistent with the current indirect measurements that expect $\gamma$ to lie between $41^{\circ}$ and $75^{\circ}$ \cite{Charles:2004jd,Bona:2005vz}. Note that the intrinsic theoretical uncertainty associated with possible U-spin breaking effects is expected to be very small and that U-spin symmetry is the only assumption that is made in this approach. As higher statistics is being accumulated by the $B$ factories, the statistical uncertainties will become even smaller. 

%DS At the moment the difference between the four values of $\gamma$ extracted from the four subsets is not very meaningful due to large uncertainties (Table~\ref{tab:fits}). When all branching ratios and $CP$ asymmetries in Tables~\ref{V0P+data}$-$\ref{P0V+data} are experimentally determined with high accuracy, U-spin approach will be capable of extracting $\gamma$ with high-precision from all four subsets of charged charmless $B$ decays. We expect that spread in $\gamma$ values to be small and indicative of the size of U-spin breaking effects. Alternatively, in case of the large spread one may speculate about possible New Physics scenarios that affect four subsets in a substantially different manner. 

\section*{ACKNOWLEDGMENTS}
We thank J.~G.~Smith for helpful discussions. This research was supported in part by DOE contract Nos.DE-FG02-04ER41291(BNL).

\pagebreak

\section*{APPENDIX A:  EXPERIMENTAL DATA}

\begin{table*}[h]
\renewcommand{\arraystretch}{1.3}
\scriptsize
\caption{$V^0 P^+$: Experimental branching ratios of charged $B$ meson decays to $V^0 P^+$. $CP$-averaged branching ratios are quoted in units of $10^{-6}$. Numbers in parentheses are upper bounds at 90 \% c.l.  
References are given in square brackets.  Additional lines, if any, 
give the direct $CP$ asymmetries ${\cal  A}_{CP}$ (second line).  The error in the average includes the scale factor $S$ when this number is shown in parentheses.
\label{V0P+data}}
\begin{ruledtabular}
\begin{tabular}{llllll}
 & Mode & CLEO & BaBar & Belle & Average \\ 
\hline
$B^+ \to$
    & $\ol{K}^{*0} K^+$ 
        & $0.0^{+1.3+0.6}_{-0.0-0.0} \; (<5.3)$ \cite{Jessop:2000bv}
        & $-$
        & $-$
        & $0.0^{+1.4}_{-0.0} \; (<5.3)$ \\
    &   & $-$
        & $-$
        & $-$
        & $-$ \\
    & $\rho^0 \pi^+$ 
        & $10.4^{+3.3}_{-3.4}\pm2.1$ \cite{Jessop:2000bv}
        & $8.8 \pm 1.0^{+0.6}_{-0.9}$ \cite{Aubert:2005sk}
        & $8.0^{+2.3}_{-2.0}\pm0.7$ \cite{Gordon:2002yt}
        & $8.7 \pm 1.1$ \\
    &   & $-$
        & $-0.07 \pm 0.12^{+0.03}_{-0.06}$ \cite{Aubert:2005sk} 
        & $-$
        & $-0.07 \pm 0.13$ \\
    & $\omega \pi^+$ 
        & $11.3^{+3.3}_{-2.9}\pm1.4$ \cite{Jessop:2000bv}
        & $6.1 \pm 0.7 \pm 0.4$ \cite{Aubert:2006qu}
        & $7.0\pm0.6\pm0.5$ \cite{Abe:2005qe}
        & $6.7 \pm 0.6$ \\
    &   & $-0.34 \pm 0.25 \pm 0.02$ \cite{Chen:2000hv}
        & $-0.01 \pm 0.10 \pm0.01$ \cite{Aubert:2006qu} 
        & $-0.03\pm0.09\pm 0.02$ \cite{Abe:2005qe}
        & $-0.04 \pm 0.07$ \\
    & $\phi \pi^+$
        & $< 5$ \cite{Bergfeld:1998ik}
        & $-0.04\pm0.17^{+0.03}_{-0.04}\; (<0.24)$ \cite{Aubert:2006nn}
        & $-$
        & $-0.04\pm0.17  \; (<0.24)$ \\
    &   & $-$
        & $-$
        & $-$
        & $-$ \\
\hline
$B^+ \to$
    & $K^{*0} \pi^+$
        & $7.6^{+3.5}_{-3.0}\pm1.6 \; (<16)$ \cite{Jessop:2000bv}
        & $13.5\pm1.2\pm0.7^{+0.4}_{-0.6}$ \cite{Aubert:2005ce}
        & $9.7 \pm 0.6 ^{+0.8}_{-0.9}$ \cite{Abe:2005ig}
        & $10.7 \pm 1.3~(S=1.60)$ \\
    &   & $-$
        & $0.068 \pm 0.078\pm 0.057^{+0.040}_{-0.035}$ \cite{Aubert:2005ce}
        & $-0.149 \pm 0.064\pm 0.031$ \cite{Abe:2005ig}
        & $-0.08 \pm 0.10~(S=1.74)$ \\
    & $\rho^0 K^+$
        & $8.4^{+4.0}_{-3.4}\pm1.8 \; (<17)$  \cite{Jessop:2000bv}
        & $5.1\pm0.8\pm0.4^{+0.2}_{-0.7}$ \cite{Aubert:2005ce}
        & $3.89 \pm 0.47^{+0.43}_{-0.41}$ \cite{Abe:2005ig}
        & $4.3\pm 0.5$ \\
    &   & $-$
        & $0.32 \pm 0.13 \pm 0.06^{+0.08}_{-0.05}$ \cite{Aubert:2005ce}
        & $0.30 \pm 0.11^{+0.11}_{-0.05}$ \cite{Abe:2005ig}
        & $0.31 \pm 0.10$ \\
    & $\omega K^+$
        & $3.2^{+2.4}_{-1.9} \pm 0.8 \; (<7.9)$  \cite{Jessop:2000bv}
        & $6.1 \pm 0.6 \pm 0.4$ \cite{Aubert:2006qu}
        & $8.1\pm 0.6\pm 0.5$ \cite{Abe:2005qe}
       & $6.9 \pm 0.9 \; (S=1.69)$  \\
    &   & $-$
        & $0.05 \pm 0.09 \pm0.01$ \cite{Aubert:2006qu}
        & $0.05\pm0.08\pm 0.01$ \cite{Abe:2005qe}
        & $0.05 \pm 0.06$ \\
    & $\phi K^+$
        & $5.5^{+2.1}_{-1.8} \pm 0.6$ \cite{Briere:2001ue}
        & $8.45\pm0.65\pm0.67$ \cite{E. Di Marco ICHEP talk}
        & $9.60 \pm 0.92\pm 0.71^{+0.78}_{-0.46}$ \cite{Garmash:2005ji}
        & $8.3 \pm 0.6^{~a}$ \\
    &   & $-$
        & $0.046 \pm 0.046 \pm0.017$ \cite{Aubert:2006av}
        & $0.01 \pm 0.12 \pm 0.05$ \cite{Chen:2003jf}
        & $0.03 \pm 0.04^{~b}$\\
\end{tabular}
\end{ruledtabular}
\leftline{$^a$Includes the CDF measurement of
${\cal B}(B^+ \to \phi K^+)=7.6\pm1.3\pm0.6$~\cite{Acosta:2005eu}.}
\leftline{$^b$Includes the CDF measurement of
$A_{CP}(B^+ \to \phi K^+) =-0.07\pm0.17^{+0.03}_{-0.02}$~\cite{Acosta:2005eu}.}
\end{table*}

\begin{table*}%[th]
\renewcommand{\arraystretch}{1.3}
\scriptsize
\caption{$P^0 P^+$: Same as Table \ref{V0P+data} for the $B^+ \to P^0 P^+$ decays.
\label{P0P+data}}
\begin{ruledtabular}
\begin{tabular}{llllll}
 & Mode & CLEO & BaBar & Belle & Average \\ 
\hline
$B^+ \to$
    & $\ol{K}^0 K^+ $ 
        & $<3.3$ \cite{Bornheim:2003bv}
        & $1.6\pm0.4\pm0.1$ \cite{Aubert:2006gm}
        & $1.22^{+0.33+0.13}_{-0.28-0.16}$ \cite{Abe:2006xs}
        & $1.4\pm0.3$ \\
    &   & $-$ 
        & $0.10\pm0.26\pm0.03$ \cite{Aubert:2006gm}
        & $0.13^{+0.23}_{-0.24}\pm0.02$ \cite{Abe:2006xs}
        & $0.12\pm0.18$ \\
    & $\pi^0 \pi^+ $ 
        & $4.6^{+1.8+0.6}_{-1.6-0.7}$ \cite{Bornheim:2003bv}
        & $5.1\pm0.5\pm0.3$ \cite{Aubert:2006ap}
        & $6.6\pm0.4^{+0.4}_{-0.5}$ \cite{Y Unno ICHEP talk}
        & $5.8\pm0.6 \, (S=1.39)$ \\
    &   & $-$
        & $-0.02\pm0.09\pm0.01$ \cite{Aubert:2006ap}
        & $0.07\pm0.06\pm0.01$ \cite{Y Unno ICHEP talk}
        & $0.04\pm0.05$ \\
    & $\eta \pi^+ $ 
        & $1.2^{+2.8}_{-1.2} \; (<5.7)$ \cite{Richichi:1999kj}
        & $5.1\pm0.6\pm0.3$ \cite{Aubert:2005bq}
        & $4.2\pm0.4\pm0.2$ \cite{J Dragic ICHEP talk}
        & $4.4\pm0.4  \, (S=1.13)$ \\  
    &   & $-$ 
        & $-0.13\pm0.12\pm0.01$ \cite{Aubert:2005bq}
        & $-0.23\pm0.09\pm0.02$ \cite{J Dragic ICHEP talk}
        & $-0.19\pm0.07$ \\
    & $\eta' \pi^+ $ 
        & $1.0^{+5.8}_{-1.0} \; (<12)$ \cite{Richichi:1999kj}
        & $4.0\pm0.8\pm0.4$ \cite{Aubert:2005bq}
        & $1.8^{+0.7}_{-0.6}\pm0.1$ \cite{Abe:2005na}
        & $2.6\pm0.8 \, (S=1.42)$ \\
    &   & $-$ 
        & $0.14\pm0.16\pm0.01$ \cite{Aubert:2005bq}
        & $0.20^{+0.37}_{-0.36}\pm0.04 $ \cite{Abe:2005na}
        & $0.15\pm0.15$ \\
\hline
$B^+ \to$
    & $K^0 \pi^+ $ 
        & $18.8^{+3.7+2.1}_{-3.3-1.8}$ \cite{Bornheim:2003bv}
        & $23.9\pm1.1\pm1.0$ \cite{Aubert:2006gm}
        & $22.9^{+0.8}_{-0.7}\pm1.3$ \cite{Abe:2006xs}
        & $23.1\pm1.0 $ \\
    &   & $0.18\pm0.24\pm0.02$ \cite{Chen:2000hv}
        & $-0.03\pm0.04\pm0.01$ \cite{Aubert:2006gm}
        & $0.03\pm0.03\pm0.01$ \cite{Abe:2006xs}
        & $0.01\pm0.02$ \\
    & $\pi^0 K^+$ 
        & $12.9^{+2.4+1.2}_{-2.2-1.1}$ \cite{Bornheim:2003bv}
        & $13.3\pm0.6\pm0.6$ \cite{Aubert:2006ap}
        & $12.4\pm0.5^{+0.7}_{-0.6}$ \cite{Y Unno ICHEP talk}
        & $12.8\pm0.6$ \\
    &   & $-0.29\pm0.23\pm0.02$ \cite{Chen:2000hv}
        & $0.02\pm0.04\pm0.01$ \cite{Aubert:2006ap}
        & $0.07\pm0.03\pm0.01$ \cite{Y Unno ICHEP talk}
        & $0.05\pm0.02$ \\
    & $\eta K^+$ 
        & $2.2^{+2.8}_{-2.2} \; (<6.9)$ \cite{Richichi:1999kj}
        & $3.3\pm0.6\pm0.3$ \cite{Aubert:2005bq}
        & $1.9\pm0.3^{+0.2}_{-0.1}$ \cite{J Dragic ICHEP talk} 
        & $2.2\pm0.4 \, (S=1.30)$ \\
    &   & $-$
        & $-0.20\pm0.15\pm0.01$ \cite{Aubert:2005bq}
        & $-0.39\pm0.16^\pm0.03$ \cite{J Dragic ICHEP talk} 
        & $-0.29\pm0.11$ \\
    & $\eta' K^+$ 
        & $80^{+10}_{-9}\pm7$ \cite{Richichi:1999kj}
        & $68.9\pm2.0\pm3.2$ \cite{Aubert:2005iy}
        & $69.2\pm2.2\pm3.7$ \cite{Schumann:2006bg}
        & $69.7\pm 2.8$ \\
    &   & $0.03\pm0.12\pm0.02$ \cite{Chen:2000hv}
        & $0.033\pm0.028\pm0.005$ \cite{Aubert:2005iy}
        & $0.028\pm0.028\pm0.021$ \cite{Schumann:2006bg}
        & $0.03\pm0.02$ \\
\end{tabular}
\end{ruledtabular}
\end{table*}

\begin{table*}[t]
\renewcommand{\arraystretch}{1.3}
\scriptsize
\caption{$V^0 V^+$: Same as Table \ref{V0P+data} for the $B^+ \to V^0 V^+$ decays.
\label{V0V+data}}
\begin{ruledtabular}
\begin{tabular}{llllll}
 & Mode & CLEO & BaBar & Belle & Average \\ 
\hline
$B^+ \to$
    & $\ol{K}^{*0} K^{*+}$ 
        & $<71$ \cite{Godang:2001sg,HFAG}
        & $-$
        & $-$
        & $<71$ \\
    &   & $-$
        & $-$
        & $-$
        & $-$ \\
    & $\rho^0 \rho^+$ 
        & $-$
        & $16.8\pm2.2\pm2.3$ \cite{Aubert:2006sb}
        & $31.7\pm7.1^{+3.8}_{-6.7}$ \cite{Zhang:2003up}
        & $18.2 \pm 4.4 \, (S=1.45)$ \\
    &   & $-$
        & $-0.12 \pm 0.13\pm0.10$ \cite{Aubert:2006sb} 
        & $0.00 \pm 0.22\pm0.03$ \cite{Zhang:2003up}
        & $-0.08 \pm 0.13$ \\
    & $\omega \rho^+$ 
        & $<61$ \cite{Bergfeld:1998ik}
        & $10.6\pm2.1^{+1.6}_{-1.0}$ \cite{Aubert:2006vt}
        & $-$ 
        & $10.6 \pm 2.5$ \\
    &   & $-$ 
        & $0.04 \pm 0.18 \pm0.02$ \cite{Aubert:2006vt} 
        & $-$ 
        & $0.04 \pm 0.18$ \\
    & $\phi \rho^+$
        & $<16$ \cite{Bergfeld:1998ik}
        & $-$ 
        & $-$
        & $<16$ \\
    &   & $-$
        & $-$
        & $-$
        & $-$ \\
\hline
$B^+ \to$
    & $K^{*0} \rho^+$
        & $-$
        & $9.6\pm1.7\pm1.5$ \cite{Aubert:2006fs}
        & $8.9 \pm 1.7 \pm 1.2$ \cite{Zhang:2005iz}
        & $9.2 \pm 1.5$ \\
    &   & $-$
        & $-0.01 \pm 0.16\pm 0.02$ \cite{Aubert:2006fs}
        & $-$
        & $-0.01 \pm 0.16$ \\
    & $\rho^0 K^{*+}$
        & $<74$  \cite{Godang:2001sg,HFAG}
        & $3.6\pm1.7\pm0.8 \, (<6.1)$ \cite{Aubert:2006fs}
        & $-$
        & $3.6\pm 1.9 \, (<6.1)$ \\
    &   & $-$ 
        & $0.20 ^{+0.32}_{-0.29}\pm 0.04$ \cite{Aubert:2003mm} 
        & $-$
        & $0.20 \pm 0.31$ \\
    & $\omega K^{*+}$
        & $<87$  \cite{Bergfeld:1998ik}
        & $0.6 ^{+1.4+1.1}_{-1.2-0.9} \; (<3.4)$ \cite{Aubert:2006vt}
        & $-$ 
       & $0.6 \pm 1.6 \; (<3.4)$  \\
    &   & $-$
        & $-$
        & $-$
        & $-$ \\
    & $\phi K^{*+}$
        & $10.6^{+6.4+1.8}_{-4.9-1.6}$ \cite{Briere:2001ue}
        & $12.7^{+2.2}_{-2.0}\pm1.1$ \cite{Aubert:2003mm}
        & $6.7^{+2.1+0.7}_{-1.9-1.0}$ \cite{Chen:2003jf}
        & $9.7 \pm 2.1~(S=1.34)$ \\
    &   & $-$
        & $0.16 \pm 0.17 \pm0.03$ \cite{Aubert:2003mm}
        & $-0.02 \pm 0.14 \pm 0.03$ \cite{Chen:2005zv}
        & $0.05 \pm 0.11$\\
\end{tabular}
\end{ruledtabular}
\end{table*}

\begin{table*}[t]
\renewcommand{\arraystretch}{1.3}
\scriptsize
\caption{$P^0 V^+$: Same as Table \ref{V0P+data} for the $B^+ \to P^0 V^+$ decays.
\label{P0V+data}}
\begin{ruledtabular}
\begin{tabular}{llllll}
 & Mode & CLEO & BaBar & Belle & Avg. \\ 
\hline
$B^+ \to$
    & $ \ol{K}^0 K^{*+}$ 
        & $-$ 
        & $-$ 
        & $-$
        & $-$ \\
    &   & $-$
        & $-$
        & $-$
        & $-$ \\
    & $ \pi^0 \rho^+$ 
        & $<43$ \cite{Jessop:2000bv}
        & $10.0 \pm 1.4 \pm 0.9$ \cite{Aubert:2005ai} 
        & $13.2 \pm 2.3^{+1.4}_{-1.9}$ \cite{Zhang} 
        & $10.8 \pm 1.5$ \\
    &   & $-$
        & $-0.01 \pm 0.13 \pm 0.02$ \cite{Aubert:2005ai} 
        & $0.06 \pm 0.19^{+0.04}_{-0.06}$ \cite{Zhang} 
        & $0.01 \pm 0.11$ \\
    & $ \eta \rho^+$ 
        & $4.8^{+5.2}_{-3.8} \; (<15)$ \cite{Richichi:1999kj}
        & $8.4 \pm 1.9 \pm 1.1$ \cite{Aubert:2005bq}
        & $4.1^{+1.4}_{-1.3}\pm 0.3$ \cite{Wang:2006cb}
        & $5.3\pm1.3 \, (S=1.17)$  \\
    &   & $-$
        & $0.02 \pm 0.18 \pm 0.02$ \cite{Aubert:2005bq}
        & $-0.04^{+0.34}_{-0.32}\pm0.01$ \cite{Wang:2006cb}
        & $0.01 \pm 0.16$ \\
    & $ \eta' \rho^+$ 
        & $11.2^{+11.9}_{-7.0} \; (<33)$ \cite{Richichi:1999kj}
        & $8.7^{+3.1+2.3}_{-2.8-1.3} \; (<22)$ \cite{Aubert:2006as} 
        & $-$
        & $9.1\pm3.2 \; (<22)$  \\
    &   & $-$
        & $-0.04\pm0.28\pm0.02$ \cite{Aubert:2006as}
        & $-$
        & $-0.04\pm0.28$ \\
\hline
$B^+ \to$
    & $K^0 \rho^+ $
        & $<48$ \cite{Asner:1996hc}
        & $-$
        & $-$
        & $<48$ \\
    &   & $-$
        & $-$
        & $-$
        & $-$ \\
    & $\pi^0 K^{*+} $
        & $7.1^{+11.4}_{-7.1}\pm1.0 \; (<31)$ \cite{Jessop:2000bv}
        & $6.9 \pm 2.0 \pm 1.3$  \cite{Aubert:2005cp}
        & $-$
        & $6.9 \pm 2.3$ \\
    &   & $-$
        & $0.04 \pm 0.29 \pm 0.05$ \cite{Aubert:2005cp}
        & $-$
        & $0.04 \pm 0.29$ \\
    & $\eta K^{*+} $
        & $26.4^{+9.6}_{-8.2}\pm3.3$ \cite{Richichi:1999kj}
        & $18.9 \pm 1.8 \pm 1.3$ \cite{Aubert:2006fj}
        & $19.7^{+2.0}_{-1.9}\pm1.4$ \cite{Wang:2006cb}
        & $19.5 \pm 1.6$ \\
    &   & $-$
        & $0.01 \pm 0.08 \pm 0.02$ \cite{Aubert:2006fj}
        & $0.03\pm0.10\pm 0.01$ \cite{Wang:2006cb}
        & $0.02 \pm 0.06$ \\
    & $\eta' K^{*+} $
        & $11.1^{+12.7}_{-8.0} \; (<35)$ \cite{Richichi:1999kj}
        & $4.9^{+1.9}_{-1.7} \pm 0.8 $ \cite{Aubert:2006as}
        & $<90$ \cite{Aihara2003}
        & $5.2\pm1.9$ \\
    &   & $-$
        & $0.30^{+0.33}_{-0.37} \pm 0.02 $ \cite{Aubert:2006as}
        & $-$
        & $0.30\pm0.35$ \\
\end{tabular}
\end{ruledtabular}
\end{table*}


\begin{thebibliography}{99}
\bibitem{angles}  
To remind, there are two alternative notations for the CKM angles, $(\beta,\alpha,\gamma)\equiv(\phi_1,\phi_2,\phi_3)$.
\bibitem{tree_foot}
Since both the u and the c piece in the $H_{eff}$ also
involve pieces of the t-quark, use of tree and penguin 
terminology for these, which is sometimes done, may be misleading.
\bibitem{units}
All branching ratios in this paper are in units of $10^{-6}$.
\end{thebibliography}
\end{document}